\documentclass[referee]{aastex62}

\RequirePackage{xspace}
\RequirePackage{amsmath}
\RequirePackage{amsfonts}
\RequirePackage{amssymb}

\def\ifundefined#1{\expandafter\ifx\csname#1\endcsname\relax}

\def\la{\mathrel{\hbox{\rlap{\hbox{\lower4pt\hbox{$\sim$}}}\hbox{$<$}}}}
\def\ga{\mathrel{\hbox{\rlap{\hbox{\lower4pt\hbox{$\sim$}}}\hbox{$>$}}}}

\newcommand{\be}{\begin{equation}}
\newcommand{\ee}{\end{equation}}

\newcommand{\bea}{\begin{eqnarray}}
\newcommand{\eea}{\end{eqnarray}}

\ifundefined{ensuremath}\def\ensuremath#1{\relax\ifmmode{#1}}
\else${#1}$\fi\else\relax\fi
\ifundefined{nuc}\def\nuc#1#2{\relax\ifmmode{}^{#1}{\protect\text{#2}}
\else${}^{#1}$#2\fi}\else\relax\fi

\newcommand{\gcm}{g~cm$^{-3}$\xspace}
\newcommand{\kmps}{\ensuremath{\text{km}~\text{s}^{-1}}\xspace}

\def\ang{\ensuremath{\text{\AA}}\xspace}

\newcommand{\phx}{\texttt{PHOENIX}\xspace}

\graphicspath{{./}}

\makeatletter

\newcommand\Autoref[1]{\@first@ref#1,@}
\def\@throw@dot#1.#2@{#1}\def\@set@refname#1{    \edef\@tmp{\getrefbykeydefault{#1}{anchor}{}}    \def\@refname{\@nameuse{\expandafter\@throw@dot\@tmp.@autorefname}s}}
\def\@first@ref#1,#2{  \ifx#2@\autoref{#1}\let\@nextref\@gobble  \else    \@set@refname{#1}    \@refname~\ref{#1}    \let\@nextref\@next@ref  \fi  \@nextref#2}
\def\@next@ref#1,#2{   \ifx#2@ and~\ref{#1}\let\@nextref\@gobble   \else, \ref{#1}   \fi   \@nextref#2}

\makeatother

\received{\null}
\revised{\null}
\accepted{\null}

\submitjournal{ApJ}

\shorttitle{SN~2012fr}
\shortauthors{Cain et al.}

\begin{document}

\title{Investigating the Unusual Spectroscopic Time-Evolution in SN~2012fr\footnote{This paper includes data gathered with the 6.5 m Magellan  Baade Telescope, located at Las Campanas Observatory, Chile.}
}

\correspondingauthor{E.~Baron}
\email{baron@ou.edu}

\author[0000-0001-9420-7384]{Christopher~Cain}
\affiliation{Azusa Pacific University 
901 E Alosta Ave, Azusa, California, 91702, USA}
\affiliation{University of Oklahoma 
440 W. Brooks, Rm 100, Norman, Oklahoma, 73019, USA}
\affiliation{Department of Physics and Astronomy, University of California, Riverside, CA 92521, USA}

\author[0000-0001-5393-1608]{E.~Baron}
\affiliation{University of Oklahoma 
440 W. Brooks, Rm 100, Norman, Oklahoma, 73019, USA}
\affiliation{Hamburger Sternwarte, Gojenbergsweg 112, 21029 Hamburg, Germany}
\affiliation{Visiting Astronomer, Department of Physics and Astronomy, Aarhus University, Ny Munkegade 120, DK-8000 Aarhus C, Denmark.}

\author[0000-0003-2734-0796]{M.~M.~Phillips}
\affiliation{Las Campanas Observatory, Carnegie Observatories\\
Casilla 601, La Serena, Chile}

\author[0000-0001-6293-9062]{Carlos~Contreras}
\affiliation{Las Campanas Observatory, Carnegie Observatories, Casilla 601, La Serena, Chile}
\affiliation{Department of Physics and Astronomy, Aarhus University, Ny Munkegade 120, DK-8000 Aarhus C, Denmark.}

\author[0000-0002-5221-7557]{Chris Ashall}
\affiliation{Department of Physics, Florida State University, Tallahassee, FL 32306, USA}

\author[0000-0002-5571-1833]{Maximilian~D.~Stritzinger}
\affiliation{Department of Physics and Astronomy, Aarhus University, Ny Munkegade 120, DK-8000 Aarhus C, Denmark.}

\author[0000-0003-4625-6629]{Christopher~R.~Burns}
\affiliation{Observatories of the Carnegie Institution for
 Science, 813 Santa Barbara St., Pasadena, CA 91101, USA}

\author[0000-0001-6806-0673]{Anthony~L.~Piro}
\affiliation{Observatories of the Carnegie Institution for
 Science, 813 Santa Barbara St., Pasadena, CA 91101, USA}

\author[0000-0003-1039-2928]{Eric~Y.~Hsiao}
\affiliation{Department of Physics, Florida State University, Tallahassee, FL 32306, USA}
\affiliation{Las Campanas Observatory, Carnegie Observatories, Casilla 601, La Serena, Chile}
\affiliation{Department of Physics and Astronomy, Aarhus University, Ny Munkegade 120, DK-8000 Aarhus C, Denmark.}

\author[0000-0002-4338-6586]{P.~Hoeflich}
\affiliation{Department of Physics, Florida State University, Tallahassee, FL 32306, USA}

\author[0000-0002-6650-694X]{Kevin Krisciunas}
\affiliation{George P. and Cynthia Woods Mitchell Institute for Fundamental Physics and Astronomy, 
Department of Physics and Astronomy, Texas A\&M University, College Station, TX 77843, USA}

\author[0000-0002-8102-181X]{Nicholas~B.~Suntzeff}
\affiliation{George P. and Cynthia Woods Mitchell Institute for Fundamental Physics and Astronomy, 
Department of Physics and Astronomy, Texas A\&M University, College Station, TX 77843, USA}

\begin{abstract}

The type~Ia supernova (SN) 2012fr displayed an unusual combination of its \ion{Si}{2}
$\lambda\lambda$5972, 6355 features. This includes the ratio of their
pseudo equivalent widths, placing it  at the border of  the Shallow
Silicon (SS) and Core Normal (CN) spectral subtype in the Branch
diagram, while the \ion{Si}{2}$\lambda$6355 expansion velocities
places it as a High-Velocity (HV) object in the Wang et al. spectral
type that most interestingly evolves slowly, placing  it in  the  Low
Velocity Gradient (LVG) typing of Benetti et al.
Only 5\% of SNe~Ia are HV and located in the SS+CN portion of the
Branch diagram and less than 10\% of SNe~Ia are both HV and LVG.
These features point towards  SN~2012fr being quite unusual, similar
in many ways to the peculiar SN 2000cx.
We modeled the spectral evolution of SN 2012fr to see if we could gain some insight into its evolutionary behavior.  
We use the parameterized radiative transfer code SYNOW to probe the abundance stratification of SN~2012fr at pre-maximum, maximum, and post-maximum light epochs.
We also use a grid of W7 models in the radiative transfer code \phx to probe the effect of different density structures on the formation of the \ion{Si}{2} $\lambda 6355$ absorption feature at post-maximum epochs.
We find that the unusual features observed in SN 2012fr are likely due to a shell-like density enhancement in the outer ejecta.  
We comment on possible reasons for atypical \ion{Ca}{2} absorption features, and suggest that they are related to the \ion{Si}{2} features.  

\end{abstract}

\section{Introduction} \label{sec:intro}

Type Ia Supernovae (SNe Ia) are precise distance indicators,
and are thus useful for measuring the expansion of the universe~\citep{brancharaa98}.  
Defining features of SNe Ia include (i) the relationship
between their light curve width and luminosity~\citep{philm15}, (ii)
their peak luminosity-color relationship~\citep{tripp98}, (iii) their
high degree of 
spectroscopic homogeneity~\citep{blondin12}, and (iv) their peculiar progenitors~\citep{hnuw00}.  
Efforts to sub-classify SNe Ia to increase their usefulness as ``standard candles'' have been
numerous \citep{branchcomp105,XWangUV12}.  
Most classification groups are developed based on spectroscopic features that are common and/or unique to SNe Ia.  
However, a number of odd SNe Ia have proven difficult to classify, and so their study is of particular importance in understanding the effort to solve the ``second parameter problem''. 
SN 2012fr is one such object which \citet{contreras12fr18} has
recently shown is a 2000cx-like SN Ia \citep{candia_optical_2003,li00cx01}

SN 2012fr exploded in NGC 1365, in the Fornax Cluster,  and was
discovered on October 27, 2012
\citep{childress13b,jzhang12fr14,contreras12fr18}.   
The rise time to peak UVOIR bolometric luminosity (maximum light) was estimated at $16.5\pm 0.6$ days, with the explosion time occurring on October 26 \citep{contreras12fr18}. 
The light curve shape parameter was found to be $\Delta m_{15}(B) =
0.84 \pm 0.03$ \citep{contreras12fr18}, $\Delta m_{15}(B) = 0.80$~mag 
\citep[preliminary value as reported by Contreras to Childress]{childress13b},  and $\Delta m_{15}(B) = 0.85 \pm 0.05$~mag \citep{jzhang12fr14}. 
Thus, SN 2012fr is a slow decliner and should be bright.  
SN~2012fr displayed a maximum absolute B-band magnitude of $M_{B_{\text{max}}} = -19.3$~mag (no errors reported) and its light curve decline-rate was unusually shallow for an SN Ia with such a fast rise-time \citep{contreras12fr18}.   
SN~2012fr suffered little or no host-galaxy reddening \citep{contreras12fr18}, so we made no corrections for it in this work.  We did correct for the redshift of the host-galaxy, which is $z = 0.0054$\footnote{\url{https://ned.ipac.caltech.edu/}}.

SN~2012fr does not fit nicely into existing classification schemes due
to an unusual combination of spectral features \citep{childress13b,jzhang12fr14,contreras12fr18}.  
These spectroscopic oddities affected the wavelength regions that are most
critical to existing SNe Ia classification schemes.  
Thus, developing a better understanding of the physical reasons for these features in SN~2012fr is important in the larger context of SNe Ia classification.  
The study of spectroscopically unusual SNe Ia is instrumental to developing an improved understanding of the explosion mechanism \citep{bgeb14}.  
Previous spectral studies of SN~2012fr have focused on explaining spectral features by
directly analyzing spectra \citep{childress13b,jzhang12fr14}.  
We expand on this work by analyzing SN~2012fr using both conventional and non-conventional modeling techniques.  \\

This work is organized as follows.  
\S~\ref{sec:spec} discusses the spectroscopic peculiarities of SN~2012fr and summarizes previous attempts to explain them.  
\S~\ref{sec:synow} presents SYNOW fits to a time-series of spectra and describes their qualitative behavior at pre-maximum (\S\ref{subsec:premax}), maximum light (\S\ref{subsec:max}), and post-maximum (\S\ref{subsec:postmax}) phases.  
\S~\ref{sec:density} introduces our investigation of possible unusual density structures using the parameterized deflagration model W7 in \phx;  
In \S~\ref{sec:metric}, we introduce the goodness-of-fit parameters $\lambda_{\text{diff}}$ and $R_{\lambda 6355}$ and explain their usefulness in our study. \S~\ref{sec:results} presents the results of our analysis in \phx; \S~\ref{sec:discussion} discusses implications of our results.

\section{Spectroscopic Peculiarities of SN~2012\lowercase{fr}} \label{sec:spec}

The most prominent spectroscopic anomaly in SN~2012fr is the
time-evolution of its \ion{Si}{2} $\lambda 6355$ absorption line.  
The \ion{Si}{2} $\lambda 6355$ line is the defining feature of SNe~Ia. 
It emerges just after explosion and persists past maximum light for $\sim 20$ days in most cases.
Because of its uniqueness to SNe Ia, \ion{Si}{2} $\lambda 6355$ is integral
to most sub-classification schemes.  
\citet{benetti05} used the velocity evolution of the minimum of the \ion{Si}{2} $\lambda 6355$ line, along with the light curve decline rate, $\Delta m_{15}(B)$, to classify SNe Ia.  
The subclassifications in the \citet{benetti05} system are the FAINT group, the High Velocity Gradient (HVG) group, and the Low Velocity Gradient (LVG) group.  
FAINT and HVG SNe Ia have fast \ion{Si}{2} evolution, while in LVGs the \ion{Si}{2} features evolve more slowly.   
\citet{branchcomp509} and \citet{xwang09} also made use of the \ion{Si}{2} $\lambda 6355$ feature in their classification schemes. 
\citet{branchcomp509} classified SNe Ia into four groups based on the pseudo-equivalent widths (pEWs) of the \ion{Si}{2} $\lambda 5972$ and $\lambda 6355$ lines. 
These are the Core Normal (CN), Cool (CL) Broad Line (BL), and Shallow Silicon (SS) groups. \citet{xwang09} classified SNe Ia based on the velocity of the $\lambda 6355$ line, with those having \ion{Si}{2} velocities at or above $11800$~\kmps being classified as High Velocity (HV) events.  
  
\begin{figure}[ht]
\centering
\includegraphics[scale=1]{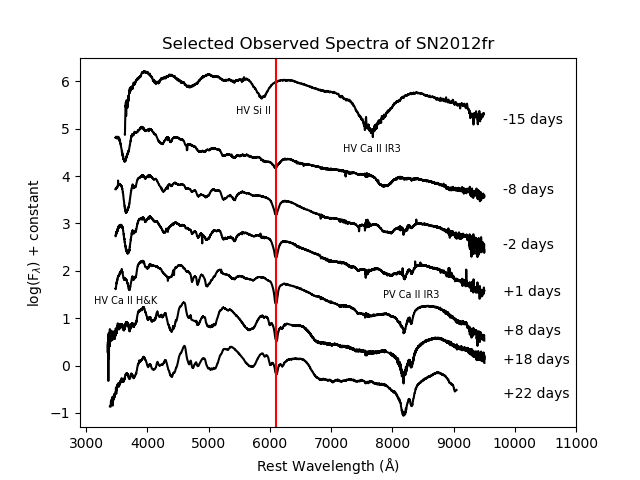}
\caption{Rest frame sample of SN~2012fr spectra ranging from day $-15$
  to day $+22$  with respect to maximum light.  The vertical red line
  through $6100 $~\ang denotes the center of the \ion{Si}{2} $\lambda
  6355$ absorption feature from day $-8$  to day $+22$,  which is the PVF component of the feature.  The HVF components of \ion{Si}{2} and \ion{Ca}{2} IR3 are labeled in the day $-15$ plot.  The HVF component of \ion{Ca}{2} H\&K can be seen at day $+8$ but not at day $+18$, and the PV \ion{Ca}{2} IR3 feature becomes prominent at abound the time HV \ion{Ca}{2} H\&K starts to weaken. Data obtained from \citet{childress13b} and \citet{jzhang12fr14}.  }
\label{fig1}
\end{figure}

 Typically, HV \ion{Si}{2} $\lambda 6355$ absorption is accompanied by HVG time-evolution.  
 As such, there is a high degree of overlap between the \citet{xwang09} HV group and the \citet{benetti05} HVG group \citep{silverman_II_12}.  
Conversely, low velocity (LV) absorption is usually accompanied by LVG features. 
However, SN~2012fr belongs to both the Benetti LVG and Wang HV groups.   
Furthermore, SN~2012fr lies on the border of the Branch CN and SS
groups, which is an unusual classification in the Branch scheme for
Wang HV events \citep{contreras12fr18,Stritz_RvsB18}.
Figure~\ref{fig1} displays a sampling of spectra ranging from day
$-15$  to day $+22$  with respect to maximum light.  All the spectra in this figure are those published in \citet{childress13b}, except the day $+12$ and day $+22$ spectra, which were published in \citet{jzhang12fr14}.  
At day $-15$  (2 days after explosion), strong \ion{Si}{2} $\lambda 6355$ absorption is centered near $5870 $~\ang ($\approx 23,000~\kmps$), which suggests a thick HV \ion{Si}{2} layer.  
This is much faster (and earlier) than usual for \ion{Si}{2} (for SN~2011fe, the \ion{Si}{2} minimum formed at $6050 \ang$).   
By day $-8$, the HV component is weak and the familiar, slower
absorption feature centered near $6100 $~\ang appears and grows
stronger, peaking in strength near maximum light.
\citet{childress13b} estimated a \ion{Si}{2} velocity of $\approx
12000$~\kmps at late times, but its temporal evolution is shallow, as
shown by the vertical red line in Figure~\ref{fig1} that
approximately bisects the minimum of the $\lambda 6355$
absorption trough from day $-8$  onward.  Furthermore, \ion{Si}{2}
$\lambda 6355$ remains visibly distinct from \ion{Fe}{2} absorption at
day $+22$  past maximum light and even up to day $+39$  according to \citet{childress13b,contreras12fr18}.

Absorption features of \ion{Ca}{2} are also characteristic of SNe Ia.  
At epochs prior to maximum light, HV \ion{Ca}{2} H\&K
tends to dominate, producing strong absorption features in the $3500-4000$~\ang wavelength region.  
These features usually become weak by maximum light and are replaced
at post-maximum phases by the Calcium Infrared Triplet (Ca IR3), which
forms in the $8000 - 8500~\ang$ region \citep{branch_pre07}.  
Typically, \ion{Ca}{2} H\&K forms in two distinct components: a
high-velocity detached component, with the detached component fading
by maximum light and a photospheric
component.
However, in SN~2012fr, High Velocity Features (HVFs) of \ion{Ca}{2} H\&K persist past maximum light.  
As seen in Figure~\ref{fig1}, strong absorption between $3500$ and
$4000$~\ang is evident between day $-8$ and day $+8$ .  
Moreover, the \ion{Ca}{2} $\lambda 8494$, $\lambda 8542$, and $\lambda 8602$ lines, which form the Ca IR3 feature, are remarkably unblended at late times (\citet{childress13b}, see their Figure 10).  
Several other SNe~Ia with similar \ion{Ca}{2} features also displayed the same kind of unusual \ion{Si}{2} behavior observed in SN~2012fr.  
The objects include SN~2000cx, SN~2006is,  SN~2009ig, and SN~2013bh \citep{contreras12fr18,Silverman13bh13}.  
It is not surprising, then, that \citet{childress13b} found that \ion{Ca}{2} and \ion{Si}{2} share a similar two-component structure and that their velocity evolution at early
and late phases are analogous.  
\citet{jzhang12fr14} noticed similar features and emphasized the predominance of intermediate mass elements (IMEs) including \ion{Si}{2} and \ion{Ca}{2} at early times.   
Both concluded from these observations that \ion{Si}{2} and \ion{Ca}{2} are likely confined within a narrow range of velocities in SN~2012fr.  
\citet{childress13b} argued that SN~2012fr has a narrowly confined shell-like distribution of IMEs persisting to late times.  
\citet{jzhang12fr14} similarly proposed that SN~2012fr is viewed at an angle from which the mass distribution is clumpy or shell-like in the outer ejecta.  
The goal of this paper is to test these assertions.

\section{SYNOW} \label{sec:synow}

We used the radiative transfer code SYNOW \citep{fisher00} to fit a
time-series of $10$ spectra ranging from day $-15$ to day $+18$  with respect to maximum light. 
SYNOW assumes homologous expansion $(v \propto r)$, spherical symmetry, a sharp photosphere with blackbody emission, and line formation by resonance scattering.  
SYNOW's main advantage is that it takes multiple scattering into account \citep{branchcomp105}. One disadvantage is that its free parameter space is large.  
SYNOW takes the photospheric velocity $v_{\text{phot}}$ and blackbody temperature $T_{\text{bb}}$ as global free parameters.  
For each ion, SYNOW takes the minimum and maximum velocities of each ion ($v_{\text{min}}$ \&  $v_{\text{max}}$), the e-folding velocity $v_{e}$ (explained below), the optical depth ($\tau$) of a reference line at $v_{\text{min}}$ for that ion, and the thermal excitation temperature $T_{\text{exec}}$.  
These parameters make SYNOW a useful tool for making line identifications and constraining ions to velocity intervals.  
However, SYNOW is not very helpful for making precise measurements because it does not give information on absolute abundances, rather only on the presence of a specific ion.
Additionally, some parameters are degenerate.  
We use it here to explore possible reasons for the behavior of \ion{Si}{2} and \ion{Ca}{2}, particularly \ion{Si}{2}, primarily on a qualitative level. 
We also use it to describe the relative evolution of other prominent ions in SN~2012fr.  \\

The global fitting parameters and line optical depths used in our fits are given in Table~\ref{tab1}.  
Although $T_{\text{exec}}$ can be set for each ion, we held it fixed at $10000$~K for all
ions up to day $+1$  and $7000$K after that, to reduce the size of parameter space, as was done in \citet{branch_pre07} and \citet{branch_post08}.  
We constrained $T_{\text{bb}}$ to $16000$ or $15000$~K at all phases (except at day $-15$, for which $T_{\text{bb}} = 7000$~K, and at day $+18$, for which $T_{\text{bb}} = 9000$~K), although previous studies have shown that this parameter is not physically meaningful, because Type Ia's are not even approximately true blackbodies and SYNOW only treats line scattering. \citep{bongard08}.  

We employ a somewhat flexible definition of $v_{\text{phot}}$ in this study.  
In some cases, we found it helpful to define the $v_{\text{phot}}$ parameter lower than the minimum velocities (described below) of photospheric ions.  
In these cases, we give both $v_{\text{phot}}$ and $v_{\text{min}}$, but
describe $v_{\text{min}}$ for the photospheric ions to be the ``true'' photospheric velocity.  Clearly, in such cases all lines are detached.
At day $+1$, $v_{\text{phot}}$ is consistent with previous fits to CN
and SS Branch-type SNe Ia near-maximum light, with the notable
exception of SN 2002cx \citep{branchcomp206}. 
We note, however, that our $+8$ day $v_{\text{phot}}$ value of $12000$~\kmps is $2000$~\kmps less than the corresponding value found by \citet{childress13b}.  
A possible reason for this discrepancy will be explained in
\S\ref{subsec:max}.  We relegate many of the figures showing detailed
SYNOW fits to \autoref{sec:appx}.

\begin{longrotatetable}
\begin{deluxetable}{ccccccccccc}
  \tabletypesize{\scriptsize}
  \tablecaption{Global SYNOW Fitting Parameters and Ion Optical Depths\label{tab1}}
\tablehead{ \colhead{Parameter} & \colhead{Day $-15$} &
\colhead{Day $-12$} & \colhead{Day $-8$} & \colhead{Day $-5$} &
\colhead{Day $-2$} & \colhead{Day $+1$} & \colhead{Day $+4$} &
\colhead{Day $+8$} & \colhead{Day $+12$} & \colhead{Day $+18$} }
                                             \startdata
                                             $v_{\text{phot}}\text{
                                               ($10^3$ km $s^{-1}$)}\tablenotemark{*}$ & $20/21$ & $14$ & $13.5$ & $13$ & $12.7/13$ & $12.8$ & $11.5/12.5$ & $11/12$ & $11/12$ & $11$\\ 
		$T_{\text{bb}} (\text{ K})$ & $7000$ & $16000$ & $16000$ & $16000$ & $15000$ & $16000$ & $16000$ & $16000$ & $15000$ & $9000$ \\
        $T_{\text{exec}} (\text{ K})$ & $1000$ & $10000$ & $10000$ & $10000$ & $10000$ & $10000$ & $7000$ & $7000$ & $7000$ & $7000$\\
        $\tau(\text{\ion{Na}{1}})$ & \nodata & \nodata & \nodata & \nodata & $0.1$ & $0.3$ & $0.3$ & $0.5$ & $0.3$ & $0.4$\\
        $\tau(\text{\ion{Mg}{2}})$ & \nodata & $2.5$ & $1.5$ & $1.3$ & $1.5$ & $1.2$ & $1.5$ & $1.8$ & $1.0$ & $1.5$\\
        $\tau(\text{\ion{Si}{2}})$ & \nodata & \nodata & $3.5$ & $3.5$ & $6.0$ & $8.0$ & $7.0$ & $6.0$ & $5.5$ & $4.0$\\
        $\tau(\text{HV \ion{Si}{2}})$ & $1.5$ & $0.6$ & $0.8$ & $0.5$ & $0.5$ & $0.8$ & $0.6$ & $0.3$ & $0.5$ & $0.3$\\
        $\tau(\text{\ion{Si}{3}})$ & $2.5$ & $1.5$ & $2.5$ & $2.5$ & $1.0$ & \nodata & \nodata & \nodata & \nodata & \nodata\\
        $\tau(\text{\ion{S}{2}})$ & $1.7$ & $1.1$ & $2.5$ & $1.9$ & $1.8$ & $1.8$ & $0.8$ & $0.4$ & $0.3$ & $0$\\
        $\tau(\text{\ion{Ca}{2}})$ & $1.0$ & $1.5$ & $2.0$ & $2.5$ & $5.0$ & $20$ & $10$ & $30$ & $50$ & $100$\\
        $\tau(\text{HV \ion{Ca}{2} H\&K})$ & $20$ & $15$ & $2.7$ & $3.0$ & $6.0$ & $3.5$ & $2.0$ & $2.0$ & $1.5$ & $2.0$\\
        $\tau(\text{\ion{Fe}{2}})$ & $0.3$ & $0.5$ & $0.4$ & $0.4$ & $0.8$ & $0.7$ & $2.0$ & $3.3$ & $5.0$ & $8.0$\\
        $\tau(\text{\ion{Fe}{3}})$ & $1.0$ & $0.8$ & $0.7$ & $0.6$ & $1.2$ & $1.5$ & $0.6$ & $0.8$ & $0.6$ & $0.4$\\
        $\tau(\text{\ion{Co}{2}})$ & \nodata & \nodata & \nodata & \nodata & $0.2$ & $1.0$ & $0.3$ & $0.5$ & $1.0$ & $2.0$\\
\enddata
\tablenotetext{*}{The notation 11/12 indicates where the
  photospheric velocity/minimum ion velocity was employed (see text).}
\end{deluxetable}    
\end{longrotatetable}

In SYNOW, the reference line optical depth of each ion as a function of velocity is approximated by  
\begin{equation}
\centering
\tau(v) = \tau(v_\text{min})e^{-(v - v_{\text{min}})/v_e},
\label{tau}
\end{equation}
where $\tau(v_{\text{min}})$ is the SYNOW input for that ion.
Smaller values of $v_{e}$ produce a higher rate of decay of $\tau$ as v increases.  The result is that $\tau$ and
$v_{e}$ are partially degenerate \citep{branchcomp206}. 
Some interesting trends are evident from the evolution of line optical
depths.  
\ion{Ca}{2} H\&K is consistently fit with distinct HV and photospheric velocity (PV) components.
The PV component is optically thin at early times and strengthens
monotonically with time, becoming quite thick by day $+18$.  
The HV component (\ion{Ca}{2} H\&K) begins optically thick, declines rapidly, then increases
again and reaches a local maximum at day $-2$.  
\ion{Si}{2} displays the same type of two-component behavior as \ion{Ca}{2}, although its HV component has a much lower $\tau$ than its PV component ($0.3 \leq \tau(\text{HV \ion{Si}{2}}) \leq 0.7$) in all later fits.   
The velocity parameters for each of the fits are given in Table~\ref{tab2}.  
The top half of the table gives $v_{\text{min}}$ and $v_{\text{max}}$ for each ion and the bottom half gives $v_{e}$.  
Ions with minimum velocities greater than the PV are referred to as ``detached.''
The behavior of $v_{\text{min}}$, $v_{\text{max}}$, and $v_{e}$ are discussed for
pre-maximum, maximum light, and post-maximum phases in \S~\ref{subsec:premax}, \S~\ref{subsec:max},
and \S~\ref{subsec:postmax}, respectively.

\begin{longrotatetable}
\begin{deluxetable}{ccccccccccc}
\tabletypesize{\scriptsize}
\tablecaption{SYNOW Velocity Parameters in $10^3$ km $s^{-1}$ \label{tab2}}
\tablehead{  \colhead{Parameter} & \colhead{Day $-15$} & \colhead{Day $-12$} & \colhead{Day $-8$} & \colhead{Day $-5$} & \colhead{Day $-2$} & \colhead{Day $+1$} & \colhead{Day $+4$} & \colhead{Day $+8$} & \colhead{Day $+12$} & \colhead{Day $+18$} } \startdata
        $v_{\text{min}},v_{\text{max}}$ (\ion{Na}{1}) & \nodata & \nodata & \nodata & \nodata & $16,19$ & $14, 16$ & $12.5, 16$ & $12, 15$ & $12, 16$ & $11, 15.5$\\
        $v_{\text{min}},v_{\text{max}}$ (\ion{Mg}{2}) & \nodata & $19,24$ & $17.5,22$ & $17,23.5$ & $15.5,20.5$ & $15.5,20$ & $13.5,17.5$ & $12.7,16.5$ & $12,16$ & $11,14$\\
        $v_{\text{min}},v_{\text{max}}$ (\ion{Si}{2}) & \nodata & \nodata & $13.5,16$ & $14,16$ & $14,15.5$ & $14,16$ & $14,15.5$ & $14,15.5$ & $13.8,15.5$ & $13.5,15$\\
        $v_{\text{min}},v_{\text{max}}$ (HV \ion{Si}{2}) & $25,37$ & $22,34$ & $16,21$ & $16,20$ & $15.5,19$ & $16,18.5$ & $15.5,18$ & $15.5,18$ & $15.5,18$ & $15,17$\\
        $v_{\text{min}},v_{\text{max}}$ (\ion{Si}{3}) & $23,28$ & $17,21$ & $14,16$ & $13,15$ & $13,15$ & \nodata & \nodata & \nodata & \nodata & \nodata\\
        $v_{\text{min}},v_{\text{max}}$ (\ion{S}{2}) & $21,24$ & $14,19$ & $13.5,16$ & $13,16.7$ & $13,16$ & $12.8,16$ & $12.5,16$ & $14,16$ & $12,14$ & \nodata \\
        $v_{\text{min}},v_{\text{max}}$ (\ion{Ca}{2}) & $21,26$ & $14,25$ & $13.5,18$ & $13,19.2$ & $13,17$ & $12.8,16.8$ & $12.5,16$ & $12,15$ & $12,15$ & $11,15$\\
        $v_{\text{min}},v_{\text{max}}$ (HV \ion{Ca}{2} H\&K) & $30,48$ & $29,35$ & $22.5,34$ & $22.7,31.7$ & $21,33$ & $20.5,29$ & $20,27$ & $17.5,25$ & $17,23$ & $18,23$\\
        $v_{\text{min}},v_{\text{max}}$ (\ion{Fe}{2}) & $21,28$ & $14,19$ & $13.5,20$ & $13,20.7$ & $12.7,19$ & $12.8,16.5$ & $11.5,15$ & $12,15$ & $12,15$ & $11,13$\\
        $v_{\text{min}},v_{\text{max}}$ (\ion{Fe}{3}) & $23,28$ & $17.5,24.5$ & $14,18$ & $15,17.5$ & $14,16$ & $12.8,17.5$ & $11.7,14.5$ & $11.5,15$ & $11.5,14.5$ & $11,13.5$\\
        $v_{\text{min}},v_{\text{max}}$ (\ion{Co}{2}) & \nodata & \nodata & \nodata & \nodata & $13,15$ & $12.8,13.5$ & $12.5,14.5$ & $12,14$ & $11.5,13.5$ & $11,13.5$\\ \hline
        $v_e$ (\ion{Na}{1}) & \nodata & \nodata & \nodata & \nodata & $1.0$ & $1.0$ & $1.0$ & $1.0$ & $8.0$ & $10$\\
        $v_e$ (\ion{Mg}{2}) & \nodata & $1.0$ & $1.0$ & $1.0$ & $1.0$ & $1.0$ & $1.0$ & $1.0$ & $1.0$ & $1.0$\\
        $v_e$ (\ion{Si}{2}) & \nodata & \nodata & $1.0$ & $1.0$ & $1.0$ & $1.0$ & $1.0$ & $1.0$ & $1.0$ & $1.0$\\
        $v_e$ (HV \ion{Si}{2}) & $4.0$ & $5.0$ & $1.0$ & $1.0$ & $1.0$ & $1.0$ & $1.0$ & $1.0$ & $1.0$ & $1.0$\\
        $v_e$ (\ion{Si}{3}) & $1.0$ & $1.0$ & $1.0$ & $1.0$ & $1.0$ & \nodata & \nodata & \nodata & \nodata & \nodata \\
        $v_e$ (\ion{S}{2}) & $1.0$ & $1.0$ & $1.0$ & $1.0$ & $1.0$ & $1.0$ & $1.0$ & $1.0$ & $1.0$ & \nodata\\
        $v_e$ (\ion{Ca}{2}) & $4.0$ & $4.0$ & $3.0$ & $3.0$ & $4.0$ & $3.0$ & $3.0$ & $4.0$ & $3.0$ & $4.0$\\
        $v_e$ (HV \ion{Ca}{2} H\&K) & $10$ & $10$ & $10$ & $7.0$ & $4.0$ & $5.0$ & $6.0$ & $5.0$ & $4.0$ & $4.0$\\
        $v_e$ (\ion{Fe}{2}) & $1.0$ & $1.0$ & $1.0$ & $1.0$ & $2.0$ & $4.0$ & $2.0$ & $2.0$ & $2.0$ & $8.0$\\
        $v_e$ (\ion{Fe}{3}) & $2.0$ & $1.0$ & $1.0$ & $1.0$ & $1.0$ & $1.0$ & $1.0$ & $1.0$ & $1.0$ & $1.0$\\
        $v_e$ (\ion{Co}{2}) & \nodata & \nodata & \nodata & \nodata & $1.0$ & $5.0$ & $1.0$ & $1.0$ & $3.0$ & $5.0$\\
\enddata
\end{deluxetable}
\end{longrotatetable}
	
\subsection{Pre-Maximum Light\label{subsec:premax}}

\begin{figure}[ht]
\centering\includegraphics[scale=1]{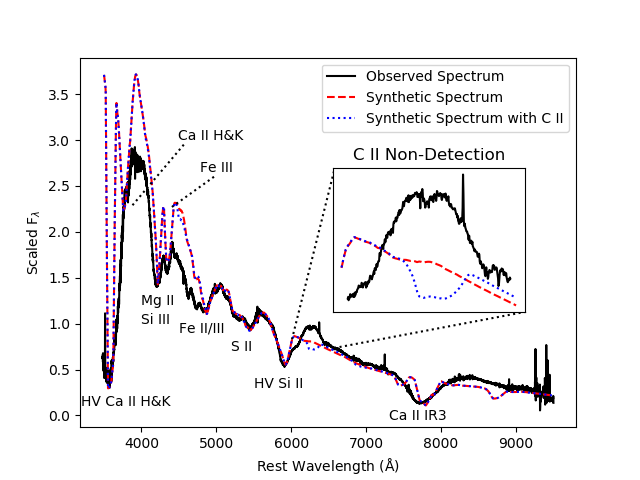}
\caption{Day $-12$ spectrum and two SYNOW fits.  The fit in red does
  not contain any \ion{C}{2}, while the dashed blue fit does.  The Ca
  IR3 fit feature is not well-blended in the fit as it is in the
  observed spectrum.  Also, adding \ion{C}{2} decreases the quality of
  the fit near $6300 $~\ang as shown in the inset.}   
\label{fig3}
\end{figure}

From day $-15$ to day $-8$, the spectra of SN~2012fr were dominated by HVFs. 
At day $-15$ and day $-12$,  HV \ion{Ca}{2} H\&K and \ion{Si}{2} absorption was
exceptionally strong. Figures~\ref{fig3} and \ref{fig:premax}  show the SYNOW fits at these times.   
At both phases, the Ca IR3 is unresolved, which is not reflected well in the fits despite large values of $v_{e}$ and $\tau(\text{HV \ion{Ca}{2}})$.  
At day $-15$, the speed of the IR3 and H\&K absorption was correctly
fit with a \ion{Ca}{2} distribution stretching up to $48000$~\kmps
 $\approx 0.16$c \citep[in agreement with][]{childress13b}, although it is unlikely that this speed is physical.  
Our fitted $\tau$ and $v_{e}$ are also large for the HV \ion{Si}{2} feature. 
\ion{O}{1} $\lambda 7773$ can account for some of the additional absorption in this region, but not all of it.  
PV \ion{Si}{2} is not identified at these phases and the PV \ion{Ca}{2} component is weak. 
Evidence for PV \ion{S}{2} is found at both times, as reflected by the
reasonable fits to the $5000 - 6000$~\ang region, where the
characteristic ``Sulfur W'' becomes distinct as the SN evolves to
maximum brightness. 
Detached \ion{Mg}{2} in the same velocity region as HV \ion{Si}{2} and \ion{Ca}{2} is also likely at day $-12$, although we find no evidence for \ion{Mg}{2} at day $-15$.  
\ion{Si}{3} $\lambda 4338.5$ may be present, although it is partially
degenerate with \ion{Mg}{2} $\lambda\lambda 4338.6$ because the two ions are used to fit the same features. 
Hence, we confirm the assertion by \citet{childress13b} that the early-time spectra are dominated by IMEs.  \\

\begin{figure}[ht]
\centering\includegraphics[scale=1]{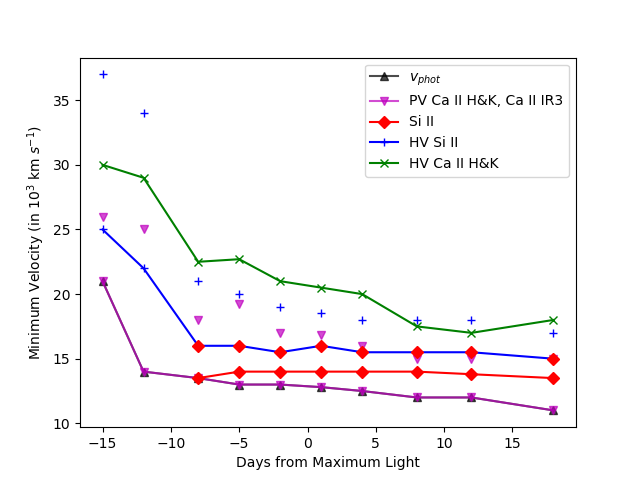}
\caption{Minimum and maximum velocities of all components of \ion{Ca}{2} and \ion{Si}{2}, as well as $v_{\text{phot}}$ vs. time WRT maximum light. The minimum ion velocity is used here for $v_{\text{phot}}$ rather than the actual parameter value (see note on Table~\ref{tab2}).  Maximum velocity points are not connected by lines in the figure, to distinguish them from the corresponding minimum velocities.  We also did not include the maximum velocity for HV \ion{Ca}{2} H\&K because it distorts the velocity scale and is not as important as the others.  Note that the minimum HV and maximum PV \ion{Si}{2} velocities are equal at day $-8$ and thereafter. } 
\label{vel_fig}
\end{figure}

We tested for \ion{C}{2} $\lambda 6580$ absorption by adding \ion{C}{2} to the day $-12$ fit with a low
optical depth of $\tau = 0.02$ in a velocity range of $v_{\text{min}},v_{\text{max}} = 17-19 \times 10^3$~\kmps.  
The fits with and without \ion{C}{2} are shown in Figure~\ref{fig3}.
We find that even optically thin \ion{C}{2} produces a worse fit near $6300$~\ang and so we did not include this feature in Tables~\ref{tab1} and~\ref{tab2}.  
Therefore, if there is any \ion{C}{2} present at early times in SN~2012fr, it is too weak to be confirmed by SYNOW in the Optical.  
However, a full search for~\ion{C}{2} would include a consideration of the NIR (see the analysis performed by \citet{marion14J15} of SN 2014J), which is beyond the scope of this work.  
This observation confirms the conclusions reached by \citet{childress13b} and \citet{jzhang12fr14}.  
\ion{Fe}{2} and \ion{Fe}{3} lines are present, although they are weak at early phases, as expected \citep{branchcomp206}.  
We note that $v_{\text{phot}}$ evolves rapidly between day $-15$ and day $-12$,
but quickly flattens out after this, as shown in Figure~\ref{vel_fig}.

Fits to the day $-8$ and day $-5$ spectra are shown in
Figure~\ref{fig:premax}  (\autoref{sec:appx}). 
By day $-8$, the HV \ion{Si}{2} feature has been largely replaced by a PV component. 
However, optically thin HV \ion{Si}{2} remains distinct well after the PV component appears.  
We differentiate between the HV \ion{Si}{2} denoted in Figures~\ref{fig:premax} and~\ref{fig3} and HV \ion{Si}{2} absorption as defined in Tables~\ref{tab1} and~\ref{tab2}.  
The former denotes visible absorption features blue-ward of $6000$~\ang and the latter denotes fitted \ion{Si}{2} with a minimum velocity greater than or equal to the maximum velocity of the PV component.  
HV \ion{Ca}{2} H\&K remains prominent surprisingly late, but PV \ion{Ca}{2}~\footnote{References to PV \ion{Ca}{2} (or just \ion{Ca}{2}) with no qualifications refer to the distribution used to fit both the H\&K and IR3 absorption features, since both lines are fit with the same set of line opacity and velocity parameters in SYNOW.  However, the detached (HV) \ion{Ca}{2} component is explicitly associated with H\&K, because it was used exclusively to fit that feature, except at day $-15$ and day $-12$.} is necessary to produce the small but distinct absorption feature near $3900 $~\ang.  
The two components remain detached from each other up to day $+4$  and possibly afterward.  
\ion{Fe}{3} is found to be detached from the photosphere, particularly at day $-5$.  
\ion{S}{2} becomes more prominent at this stage, while \ion{Mg}{2} weakens but stays detached.  \\

The detachment of \ion{Ca}{2}, \ion{Si}{2}, \ion{Mg}{2}, and even \ion{Fe}{3} at pre-maximum phases implies that SN~2012fr is highly stratified before maximum light.    
The large width of HV \ion{Si}{2} and \ion{Ca}{2} absorption features at early times suggests the presence of processed IMEs at high velocities, since it is unlikely that the progenitor of SN~2012fr would contain that much primordial \ion{Si}{2} and \ion{Ca}{2}. 
By day $-5$, this layer has become optically thin, but remains identifiable in the fits.  
We note that while the HV component of \ion{Ca}{2} remains detached from the PV component well past maximum light, there is no gap between the HV and PV components of \ion{Si}{2} at day $-8$.  Therefore, we find that the \ion{Ca}{2} distribution in SN~2012fr is discontinuous, while the \ion{Si}{2} distribution is continuous, but with a discontinuous $\tau$ gradient.  
\ion{S}{2} displays roughly the same velocity evolution as PV \ion{Ca}{2}, while \ion{Mg}{2} behaves more like the HV \ion{Si}{2} component, at least before maximum light.  
All of this suggests that the distribution of IMEs at pre-maximum phases have a complicated layered structure in which different ions dominate each layer.

\subsection{Near Maximum Light \label{subsec:max}}

Near maximum light, PV components dominate the spectra of SN~2012fr, with the exception of \ion{Ca}{2} H\&K, which continues to have a prominent HV component.  
At day $-2$  (Figure~\ref{fig3}) the HV and PV components of \ion{Si}{2} have merged, occupying the range of $14000 - 19000$~\kmps, although at different optical depths.  
As mentioned, $\tau(PV \text{\ion{Si}{2}})$ peaks at maximum light and decreases steadily afterwards.  
The first evidence for optically thin \ion{Na}{1} is found at day $-2$.  
However, the fit to the region where Na features typically form ($5700
- 6000 $~\ang) is not good at day $-2$ or $+1$, so \ion{Na}{1} may not be present at this phase (or later).     
Ca IR3 absorption is relatively weak by this time, but the Ca H\&K feature remains prominent, with distinct HV and PV \ion{Ca}{2} being needed to obtain a good fit.  
\ion{Fe}{2} becomes more prominent at this phase and PV \ion{Co}{2} appears for the first time.  \\

\begin{figure}[ht]
\centering
\includegraphics[scale=1]{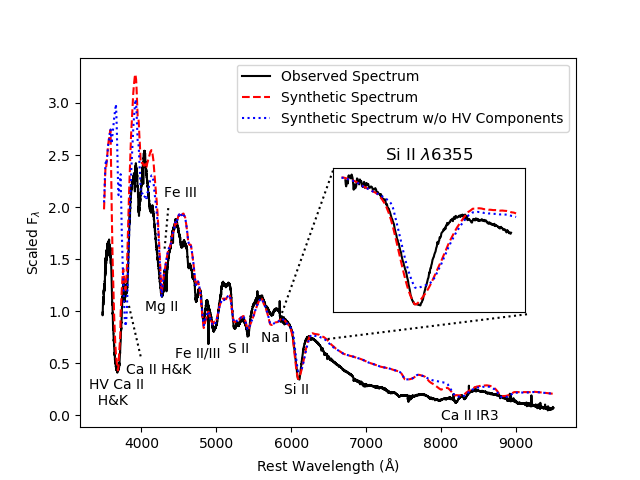}
\caption{Day $+1$ spectrum and a synthetic SYNOW fit.  The fit
    is plotted with and without the HV components of \ion{Si}{2} and
    \ion{Ca}{2} to illustrate their importance.  The \ion{Si}{2}
    $\lambda 6355$ line is enlarged in the inset to show the difference in the fits.  }
\label{fig7}
\end{figure}

As early as day $-5$ (Figure~\ref{fig3}), the PV \ion{Si}{2} distribution becomes
detached from the photosphere and the \ion{Si}{2} $\lambda 6355$ absorption
line begins to narrow after that.  
\ion{Si}{3} disappears at day $+1$, leaving a single-component Si feature with an optical depth
gradient that is discontinuous at around $15500 - 16000$~\kmps.  
All the other ions are photospheric by day $+1$  except \ion{Na}{1},
HV \ion{Ca}{2} H\&K, and \ion{Mg}{2} (Figure~\ref{fig7}).  
At day $+4$ (Figure~\ref{fig8}), the detachment of \ion{Si}{2} becomes more pronounced as the photosphere continues to recede but the minimum velocity of PV \ion{Si}{2} remains fixed at $14000$~\kmps\footnote{This could explain the aforementioned discrepancy between our value of ${v_{phot}}$ and that of \citet{childress13b} at day $+8$.  If \citet{childress13b} used the minimum of the \ion{Si}{2} line to locate the photosphere (a reasonable assumption for most SNe~Ia), then their $v_{phot}$ would correspond to our minimum \ion{Si}{2} velocity.  }. 
The \ion{Si}{2} $\lambda 6355$ line becomes progressively narrower, reflecting the decreasing size of the velocity interval in which optically thick \ion{Si}{2} is found. 
PV \ion{Ca}{2} becomes optically thick at this phase and Ca IR3 begins to strengthen. 
Absorption lines of Fe in the $4500 - 5000 $~\ang region, which become
distinct starting at day $+4$, are best fit when \ion{Fe}{2} and \ion{Fe}{3} are
placed below the minimum velocities of the photospheric ions.  This 
just reduces the effective optical depth, since SYNOW only calculates
the region between $v_{phot}$ and $v_{max}$.

The defining feature of the temporal evolution near maximum light is the detachment of \ion{Si}{2}, which becomes definite by day $-2$. 
Interestingly, the PV component of \ion{Ca}{2} (as constrained by the
PV H\&K component~\footnote{\citet{childress13b} noted that by
  contrast, the Ca IR3 lines remain at the same velocity as
  \ion{Si}{2} $\lambda6355$ at late times.  Based on the tight
  constraints of the PV \ion{Ca}{2} parameters provided by the H\&K
  feature, which suggests that \ion{Ca}{2} does evolve with the
  photosphere, we infer that the detachment of the IR3 lines is not
  the result of a radial cutoff in the \ion{Ca}{2} distribution above
  the photosphere.  Rather, the apparent high speed of the absorption
  minima could be the result of the rapidly increasing opacity of the
  lines, as explained in \citet{marion14J15} in their analysis of
  \ion{Mg}{2} in SN 2014J.  There could also be complicated Non-LTE
  (NLTE) effects that are not treated here.}) does not behave this way, remaining truly photospheric at later times.  
This is the first major divergence between the behavior of PV \ion{Ca}{2} and \ion{Si}{2}, suggesting that \ion{Ca}{2} reaches to much lower velocities than does \ion{Si}{2}.  
This explains why the PV component of \ion{Ca}{2} does not display the same LVG behavior at late times that \ion{Si}{2} does. 

\subsection{Post-Maximum Light \label{subsec:postmax}}

At post-maximum phases, the \ion{Si}{2} $\lambda 6355$ line becomes exceptionally narrow and PV \ion{Ca}{2} becomes optically thick (as evidenced by the strengthening IR3 features).  Lines of \ion{Fe}{2}, including \ion{Fe}{2} $\lambda 4924$, $\lambda 5018$, and $\lambda 5169$, also become more prominent at this stage.  
Figure~\ref{fig8} displays the fit to the day $+8$ spectrum.  
The defining feature here is the excellent fit to the Fe features in
the range $4500 - 5000 $~\ang. These lines are exceptionally well-distinguished from each other, so they constrain the \ion{Fe}{2} parameters quite well.  
Ca H\&K is still fit reasonably well in the blue, but the IR3 feature
is unresolved.  Emission near $4000$~\ang becomes much too strong starting at day $+8$, but this is likely an artifact of the blackbody approximation used in SYNOW.   
By day $+12$ (Figure~\ref{fig8}), the absorption feature in the region of \ion{Si}{2} $\lambda 5972$ broadens substantially, but by this phase the spectrum is dominated by \ion{Fe}{2} absorption, so our identification of this feature with \ion{Si}{2} $\lambda 5972$ is provisional.  
\ion{S}{2}, still faintly visible in the small absorption feature near $5400 $~\ang at day $+8$, is essentially gone by day $+12$ and entirely absent in the day $+18$ fit (Figure~\ref{fig8}).  \\

The defining feature of the day $+12$ and day $+18$ spectra is the
decreasing width of \ion{Si}{2} $\lambda 6355$.  
The absorption feature of \ion{Si}{2} $\lambda 6355$ remains distinct from strengthening \ion{Fe}{2} lines as late as day $+39$, according to \citet{childress13b}.  
The minimum velocity does finally recede slightly, to $13500$~\kmps by day $+18$, while the
$\tau(\text{\ion{Si}{2}})$ discontinuity recedes to $15000$~\kmps by that phase (Figure~\ref{vel_fig}).  
However, it still evolves much more slowly in velocity space than the photosphere does, as illustrated in Figure~\ref{vel_fig}.   
This type of \ion{Si}{2} behavior at late times was identified by \citet{bran01b} in SN 1991bg.  
The PV component of \ion{Ca}{2} in the fits does not become detached from the photosphere, but the HV component does remain identifiable and detached from the PV component.  
Therefore, the behaviors of \ion{Si}{2} and \ion{Ca}{2} continue to diverge at late times, suggesting that these ions are distributed differently at low velocities. \\

The confinement of \ion{Si}{2} to a narrow interval of velocities and the late-time distinctiveness of the \ion{Si}{2} $\lambda 6355$ line support the hypothesis that SN~2012fr may have a density enhancement after maximum light.  
But if there is a shell, it is unclear where it is located.  
At least three explanations are possible.  
First, there may be a high-velocity shell in the middle of the \ion{Si}{2} distribution.  
In this case, we would expect \ion{Ca}{2} and \ion{Si}{2} to both be detached, unless \ion{Ca}{2} is not abundant in the shell region.  
It is also possible that the narrow \ion{Si}{2} absorption is produced by a steep density profile at high velocities.  
In this case, a shell at lower speeds might explain the shallow photospheric velocity noted by \citet{contreras12fr18}.  Another possibility is that the abundance structure of SN~2012fr is sufficiently stratified to produce a structure involving multiple shells.  
The rest of this work is devoted to exploring the first two possibilities.  
Furthermore, in \S~\ref{sec:density} -~\ref{sec:results}, we investigate whether the strange behavior of SN~2012fr can be explained by only an abnormal density structure.

\section{Density Modeling in \phx} \label{sec:density}

The shell-like behavior of \ion{Si}{2} and \ion{Ca}{2} at late times
and the narrow width of the \ion{Si}{2} $\lambda 6355$ line are
independently suggestive of the presence of a shell and/or a
  steep density dropoff described above.  
\citet{jzhang12fr14} found that a value of $v_{e}$ lower than $1000$~\kmps for \ion{Si}{2} was needed to fit the maximum light \ion{Si}{2} $\lambda 6355$ feature of SN~2012fr using SYNOW.  
We obtained nearly equivalent fits with $v_{e} = 1000$~\kmps, but with an awkwardly discontinuous opacity gradient structure.  
It may be that there is actually one continuous opacity gradient with $v_{e} < 1000$~\kmps.  Either way, a possible interpretation of this result is that the density profile of SN~2012fr is steeper than most SNe, which are usually well-fitted with a single \ion{Si}{2} component with $v_e \geq 1000$~\kmps \citep{branchcomp206, branch_pre07,branch_post08}.  
This is an odd conclusion to reach, considering our earlier deduction that the region containing \ion{Si}{2} may be shell-like.  
To try to explain this phenomenon, we examined the density structure of SN~2012fr more closely.  \\

To probe the density structure of SN~2012fr, we used the parameterized deflagration model W7 \citep{nomw7} converged with the radiation transfer code \phx \citep{bbbh06}. 
We used W7 because it has a relatively simple abundance/density structure that was easy to manipulate and it fits the observed spectrum of SNe Ia relatively well near maximum light.  
Our focus was the formation of the \ion{Si}{2} $\lambda 6355$ line at post-maximum phases.  
Note that the distribution of \ion{Si}{2} in W7 is in good agreement with the velocity limits inferred from SYNOW at maximum light for SN 2012fr after being shifted $4000$~\kmps to the right, as shown in Figure~\ref{fig:si_extent}.  
Since there is no sharp photosphere in \phx, an offset in velocity space is not surprising.  
The advantage of using \phx here is that it can be used to probe the detailed physical structure of the model, not just to produce synthetic spectra.  
This allowed for direct manipulation of the density profile of W7.  
Models were converged in Local Thermal Equilibrium (LTE) for computational expediency.  
We simulated the qualitative features discussed above using three parameterized density modification procedures. 
Since W7 assumes homologous expansion, all physical manipulations were performed in velocity space.  
In each procedure, a {W7} model was calculated to provide an unmodified reference density profile $\rho_{\text{ref}}(v)$.  
This profile was piecewise fit to a set of exponential and linear functions to obtain an ``idealized'' density profile $\rho_{\text{fit}}(v)$.  \\

Next, $\rho_{\text{fit}}(v)$ was modified using three functions formulated to simulate the desired features.  
These are described in \S~\ref{subsec:d0}, \S~\ref{subsec:d1}, and \S~\ref{subsec:d2}.  
Each was applied to the fit density on a range of velocities ($v_{\text{min}}$ to $v_{\text{max}}$) such that the new fit $\rho_{\text{new}}(v)$ is given by

\begin{equation}
\label{new_def}
\large
\rho_{\text{new}}(v) = 
\begin{cases} 
\rho_{\text{mod}}(v) & v_{\text{min}} \leq v \leq v_{\text{max}}, \\ 
\rho_{\text{fit}}(v) & \text{otherwise}, 
\end{cases}
\end{equation}
where $\rho_{\text{mod}}(v)$ is the modified density function added by the procedure.  
Finally, the reference profile was multiplied at every point by the ratio of the modified and unmodified fit densities, giving 
\begin{equation}
\label{final_def}
\rho(v) = \rho_{\text{ref}}(v) \cdot \frac{\rho_{\text{new}}(v)}{\rho_{\text{fit}}(v)}.
\end{equation}

The final step was to normalize the new density to conserve the mass of the reference profile.  We integrate (using a trapezoidal approximation) to get the mass ratio $R_m$ as

\begin{equation}
\label{norm_int}
R_m = \frac{\int_{v_{lower}}^{v_{upper}}\rho_{\text{ref}}(v)v^2 \text{dv}}{\int_{v_{\text{lower}}}^{v_{upper}}\rho(v)v^2 \text{dv}},
\end{equation}
where $v_{\text{upper}}$ and $v_{\text{lower}}$ are the lower and upper velocity limits of the model. 
Models were then converged with density profile $R_m \rho(v)$.  
The advantage of this procedure is that local features of the W7 density that are not included in the idealized fit are preserved in the modified profile.  
This allows for an honest evaluation of the effects of the added features.  
We note that this method affects only the density structure of W7, it does not effect the velocity ranges where different ions are present, so the abundance structure is manipulated only indirectly.  
Future work will investigate whether a direct manipulation of the
abundance stratification of W7 or some other model can best explain
SN~2012fr.  

\subsection{Density Model: D0} \label{subsec:d0}

\begin{figure}[ht]
	\centering
        \gridline{	\fig{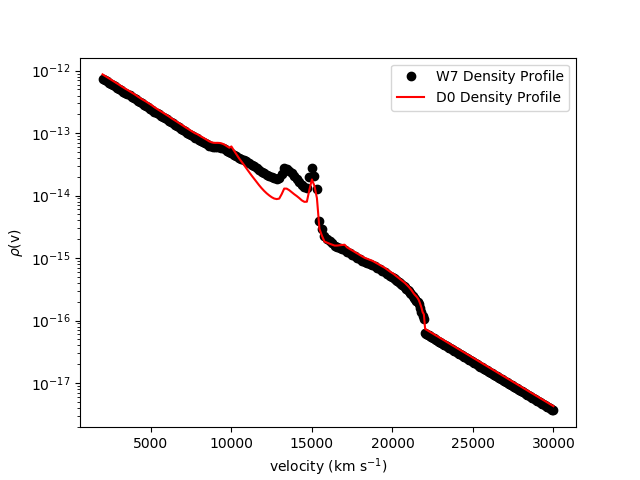}{0.45\textwidth}{(a)}
    \fig{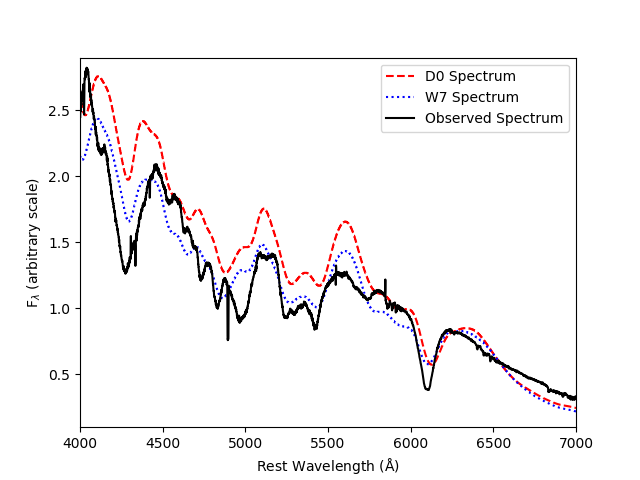}{0.45\textwidth}{(b)}
}
	\caption{W7 density profile with modified D0 profile (a) and corresponding day $+1$ fit (b) with the W7 fit with the unmodified profile included for comparison.  The input parameters for this fit are $v_{min} = 10,000$~\kmps, $v_{max} = 17,000$~\kmps, and $v_e = 1,000$~\kmps.  Note that local features of the profile, such as the irregularity centered near $14,000$~\kmps, are preserved in D0.  The flux spectra are normalized at $6260 $~\ang to make the $\lambda 6355$ line easy to compare, which is why the fit to the blue of $5800$~\ang is vertically offset.  
}
\label{fig12}
\end{figure}

Our first procedure, denoted D0, steepens the decay rate of $\rho(v)$
in the velocity region $v_{\text{min}} \le v \le v_{\text{max}}$,
where $v_{\text{min}}$ and $v_{\text{max}}$ are parameters.  
The steepened profile is an exponential function with decay rate specified by $v_{e}$ (also a free parameter), as in SYNOW.  
The modified density function is 
\begin{equation}
\label{D0}
\rho_{\text{mod}}(v) = Ae^{-v/v_{e}} + B,
\end{equation}
where $A$ and $B$ both have units of density and are chosen so that $\rho_{\text{mod}}(v_{\text{min}})=\rho_{\text{fit}}(v_{\text{min}})$ and $\rho_{\text{mod}}(v_{\text{max}}) = \rho_{\text{fit}}(v_{\text{max}})$.  
This restriction ensures that $\rho_{\text{new}}(v)$ is continuous, but requires the shift parameter $B$.  
A D0 profile with parameters $v_{\text{min}} = 10000$~\kmps, $v_{\text{max}} = 17000$~\kmps and $v_{e} = 1000$~\kmps is displayed in Figure~\ref{fig12} (left) with the corresponding day $+1$ fit to SN~2012fr (right).  
We observe that the endpoints of the modified profile are not precisely at $10000$~\kmps and $17000$~\kmps.  
This is because the density grid in velocity space only contains $256$ bins, so a binning error of $\approx 100$~\kmps is introduced.  
The local features of the W7 profile are ``warped'' to match the change to the global features.  \\
 
We held $v_{e}$ at $1000$~\kmps to reduce the size of parameter space, and varied $v_{\text{min}}$ and $v_{\text{max}}$ over a reasonable range.  
This grid-based approach allowed us to explore the dependence of the \ion{Si}{2} $\lambda 6355$ line on the D0 parameters. 
While D0 was able to reproduce the shape of the \ion{Si}{2} absorption profile reasonably well, it placed the minimum of the line too far to the red.    
This implies that in D0, the line forming region of \ion{Si}{2} is too far
inside the ejecta and thus, at  velocities that are too low.  

\subsection{Density Model: D1} \label{subsec:d1}

\begin{figure}[ht]
	\centering
\gridline{	\fig{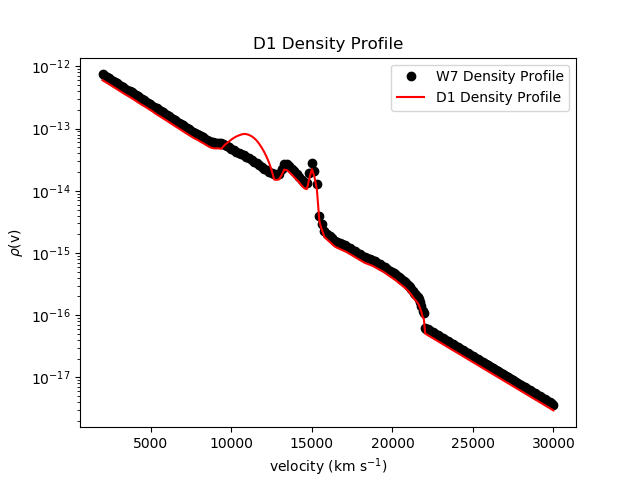}{0.45\textwidth}{(a)}
    \fig{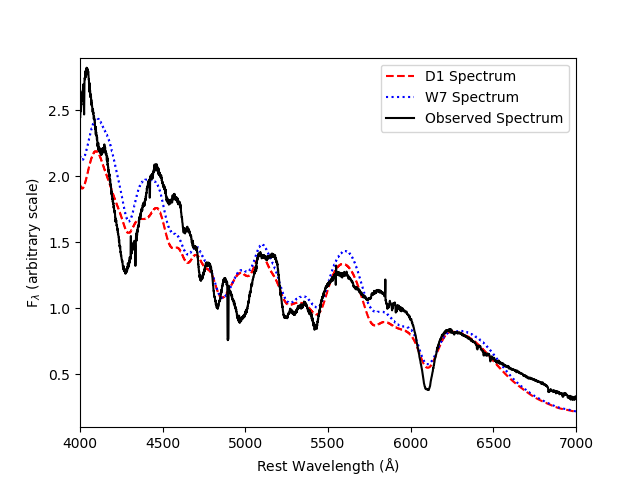}{0.45\textwidth}{(b)}
}
	\caption{W7 Density Profile with modified D1 profile (a) and corresponding day $+1$ fit (b) with the W7 fit with the unmodified profile included for comparison.  The input parameters for this fit are $v_{c} = 11,000$~\kmps, $p = 2$, and $w = 3,000$~\kmps.  }
	\label{fig13}
\end{figure}

Our second procedure, denoted D1, adds a shell-like density enhancement centered at a variable velocity $v_{c}$ with width and thickness parameters w and p.  
The shell is modeled by a ``tilted Gaussian'' function of the form

\begin{equation}
\label{D1}
\rho_{\text{mod}}(v) = (p + 1)*\rho_{\text{fit}}(v_c)e^{-(v - v_c)^2/a} + bv + c,
\end{equation}
where $a$, $b$, and $c$ are free parameters chosen so that $\rho_{mod}(v)$
fits the points $(v_{c}, (p+1)*\rho_{\text{fit}}(v_{c}))$, $(v_{c} -0.5w, \rho_{\text{fit}}(v_{c} - 0.5w))$, and $(v_{c} + 0.5w, \rho_{\text{fit}}(v_{c} + 0.5w))$.  
The result is a shell centered near $v_{c}$ with approximate width $w$ and peak
density close to $(p + 1)*\rho_{\text{fit}}(v_c)$. 
D1 reproduces the shell-like structure hypothesized in SN~2012fr with
variable dimensions and position in velocity space.  
An example D1 profile with parameters $v_{c} = 11000$~\kmps, $w = 3000$~\kmps, and $p = 2$, which clearly shows the effect of D1 on W7 and provides a reasonable fit, and the corresponding day $+1$ fit are shown in Figure~\ref{fig13}. 
The fit to the \ion{Si}{2} $\lambda 6355$ line is better than in D0 in terms of position, but the overall shape of the absorption profile is not quite the same as the observed feature.  
Also, the prominent absorption feature near $5750$~\ang (a blend
\ion{Si}{2} $\lambda 5972$, \ion{Na}{1}, and \ion{Fe}{2} absorption) that is nearly absent in SN~2012fr is much stronger in D1 than in D0, at least for these parameters.  
D0 and D1 produce partially inverse effects when applied to the same velocity regions.  
D0 subtracts mass between $v_{\text{min}}$ and $v_{\text{max}}$, while D1 adds mass between $v_c - 0.5w$ and $v_c + 0.5w$.  
The fits in Figures~\ref{fig12} and~\ref{fig13} suggest that D0 does a better job of reproducing the shape of the SN~2012fr \ion{Si}{2} absorption profile, but D1 correctly places the line in velocity space.  Thus, they account for different aspects of the observed profile.

\subsection{Density Model: D2} \label{subsec:d2}

\begin{figure}[ht]
	\centering
\gridline{	\fig{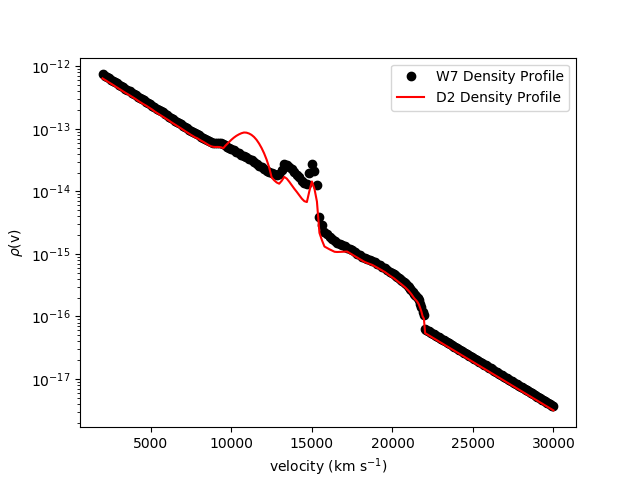}{0.45\textwidth}{(a)}
    \fig{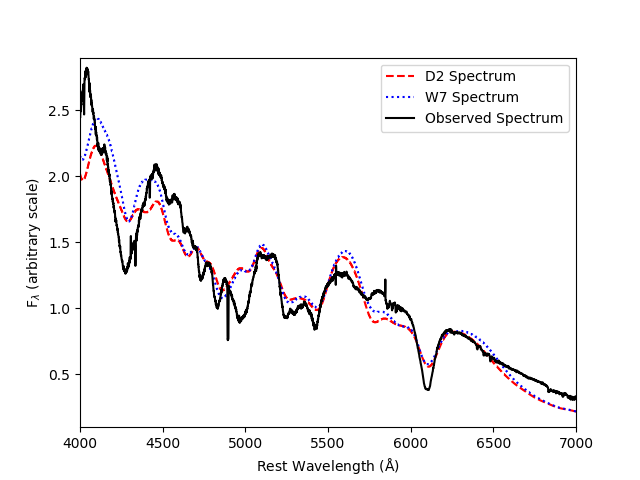}{0.45\textwidth}{(b)}
}
	\caption{W7 density profile with D2 profile (a) and corresponding day $+1$ fit (b) with the W7 fit with the unmodified profile included for comparison.  The input parameters for this fit are $v_{c} = 11,000$~\kmps, $p = 2$, $w = 3,000$~\kmps, and $v_{u} = 17,000$~\kmps.  Note that D2 adds mass at lower velocities and removes mass at higher velocities, so the effect of normalization on the rest of the profile is less than in either D0 or D1.  }
	\label{fig14}
\end{figure}

To determine how these features interact with each other, we parameterized a third procedure (denoted D2) that combines the effects of D0 and D1 by producing a shell at lower velocities and a steep density profile at higher velocities.  
D2 fits Equation~\ref{D1} with a set of D1 parameters $v_{c}$, $w$, and $p$, then fits Equation~\ref{D0} to $v_{\text{min}} = v_{c} + 0.5w$ and $v_{\text{max}} = v_u$, where
$v_u$ is a free parameter specifying the upper velocity limit for Equation~\ref{D0}. 
We fixed $v_{e} = 1000$~\kmps to reduce the size of parameter space.  
A D2 profile with parameters $v_{c} = 11000$~\kmps, $w = 3000$~\kmps, $p
= 2$ and $v_u = 17000$~\kmps is shown in Figure~\ref{fig14} (left)
and corresponding day $+1$ fit (right).  
The fit to the \ion{Si}{2} $\lambda 6355$ line is similar to Figure~\ref{fig13}.  
This is not surprising, since the D1 component of the modification overlaps the W7 \ion{Si}{2} distribution more than does the D0 component for these parameters.  
When $v_{c}$ is reduced with $v_u$ held fixed, the effect of D0 becomes
more pronounced as more mass is removed from the \ion{Si}{2} distribution.   

W7 contains Si between $9000$~\kmps and $16000$~\kmps \citep{bongard08}, so we focused our analysis on that velocity region.  
In D1 and D2, we held p to either $1$, $2$ or $3$ and w to either $2000$~\kmps or $3000$~\kmps.  We varied the other parameters to sufficiently cover the $9000 - 16000$~\kmps range.  
This process was repeated at selected phases to produce a time-series of grids.  
Because of the relatively small number of free parameters, we were able to accomplish a comprehensive analysis by converging on the order of $1000$ models.  
To quantitatively analyze the effect of these parameters on the \ion{Si}{2} $\lambda 6355$ line, we developed goodness-of-fit parameters described in the next section.  

\vspace{1cm}

\section{$\lambda_{\text{\lowercase{diff}}}$ \lowercase{and} $R_{\lambda 6355}$} \label{sec:metric}

To capture the behavior of the \ion{Si}{2}$\lambda 6355$ line quantitatively, we developed a two-parameter metric for evaluating its goodness-of-fit to observed spectra.  
We defined parameters to (1) measure the relative minima of the synthetic and observed line and (2) capture their relative absorption strengths. 
First, we define

\begin{equation}
\large
\label{lambda_diff}
\lambda_{\text{diff}} = \lambda^\text{obs}_{\text{min}} - \lambda^\text{syn}_{\text{min}},
\end{equation}
where $ \lambda^\text{obs}_{\text{min}}$ is the wavelength that
minimizes the observed flux, $F^{\text{obs}}_{\lambda}$ between $5970
$~\ang and $6220 $~\ang and $\lambda^\text{syn}_{\text{min}}$ is defined equivalently.  
Because the minimum of the \ion{Si}{2} $\lambda 6355$ line depends on
the average speed of the \ion{Si}{2} distribution, $\lambda_{\text{diff}}$ is a measure of the
difference of the average speeds of \ion{Si}{2} in the synthetic and observed spectra.  
Positive values of $\lambda_{\text{diff}}$ indicate faster synthetic absorption.  
Note that $\lambda_{\text{diff}}$ is independent how the spectra are normalized.

To measure the relative absorption strength of the lines, we first define
the continuum flux as the straight line that goes through the two
points at $(F^X_\lambda(5970),F^X_\lambda(6220))$, where the superscript $X$ refers to
synthetic or observed flux. 
Thus,

\begin{equation}
F^X_{\text{cont}}(\lambda) \equiv \left(\frac{F^X_{\lambda}(6220~\ang) - F^X_{\lambda}(5970~\ang)}{250\text{\AA}}\right)(\lambda - 5970~\ang) + F^X_{\lambda}(5970~\ang).
\end{equation}

Then we define the ratio
\begin{equation}
R_{\lambda 6355} = \frac{\int_{5970~\ang}^{6220~\ang} \left(F^{\text{syn}}_{\text{cont}}(\lambda) - F^{\text{syn}}_{\lambda}(\lambda)\right)\,d\lambda}{\int_{5970~\ang}^{6220~\ang} \left(F^{\text{obs}}_{\text{cont}}(\lambda) - F^{\text{obs}}_{\lambda}(\lambda)\right)d\lambda},
\end{equation}
where the subscripts syn and obs refer to the synthetic and observed flux, as in Equation~\ref{lambda_diff}.  
$R_{\lambda 6355}$ is conceptually similar to the ratio of the pseudo-equivalent line widths (pEWs) of \ion{Si}{2}
$\lambda 6355$ in the synthetic and observed spectra, with two important differences.  First, the flux integrals are not normalized with respect to the continuum flux, so that differences in absorption strength can be compared.  
Note that this means that $R_{\lambda 6355}$ is sensitive to flux normalization, unlike the pEW.  
We account for this by normalizing all the spectra in the same place ($6260$~\ang).  
Second, the wavelength interval used to calculate $R_{\lambda 6355}$ is held fixed at $5970 - 6220$~\ang, rather than being selected for each individual spectrum based on the particular shape of the P-Cygni line \citep{nordin11}.  
Furthermore, the pEW is usually not considered when the pseudo-continuum becomes no longer accurately measurable due to Fe lines \citep{childress13b}.  
However, $R_{\lambda 6355}$ retains its usefulness as a figure of merit because it measures how ``washed out'' the \ion{Si}{2} line is by Fe features at these times.  
Moreover, $R_{\lambda 6355}$ accounts for differences in line width because the wavelength limits of the integral are independent of differences in pEW.  \\

\begin{figure}[ht]
\centering
\includegraphics[scale=0.85]{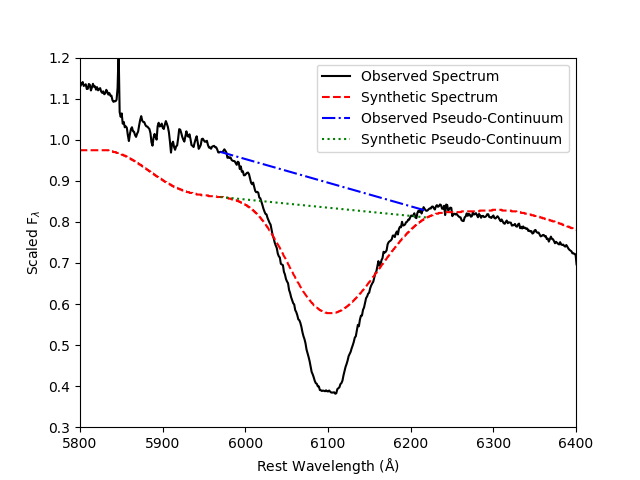}
\caption{A W7 fit to the \ion{Si}{2} $\lambda 6355$ line at day $+1$.
The dashed blue line denotes the OPC, and the dashed green line the SPC.  At day $+1$, $\lambda_{\text{diff}} = +5.55 $~\ang and $R_{\lambda 6355} = 0.564$.}

\label{fig15}
\end{figure}

Figure~\ref{fig15} displays W7 fits to the \ion{Si}{2} $\lambda 6355$
line at day $+1$ and day $+18$.  In each plot, the solid blue line is the observed pseudo-continuum (OPC) from $5970 - 6220$~\ang and the dashed blue line is the synthetic pseudo-continuum (SPC).  
In the day $+1$ fit (Figure~\ref{fig15}), the OPC clearly approximates the continuum of the observed spectrum.  
The SPC is much flatter, but it still clearly denotes the area carved out by the synthetic \ion{Si}{2} absorption.

We found that $\lambda_{\text{diff}}$ was slightly larger than $0$ in the unmodified fits and increased past maximum light.  
This systematic offset may be due to the fact that \ion{Si}{2} $\lambda 6355$ is slightly too fast in W7 \citep{bbbh06}. 
Also, the presence of Fe features in that region at late times can have the effect of masking the true position of the \ion{Si}{2} $\lambda 6355$ line.  
The doublet nature of the \ion{Si}{2} $\lambda 6355$ line can also result in the flux minimum being offset from the true center by roughly $\pm 5-10 $~\ang.   
We also found that the synthetic \ion{Si}{2} $\lambda 6355$ absorption profile in W7 was consistently too shallow.  
A perfect fit would produce $R_{\lambda 6355} = 1$, but we found that $0.2 \leq R_{\lambda 6355} \leq 0.65$ in nearly all our fits.  
Still, higher values of $R_{\lambda 6355}$ consistently correspond to better fits at fixed phases.  
We therefore focused on using $\lambda_{\text{diff}}$ and $R_{\lambda 6355}$ to make differential comparisons of the effects of changing inputs in D0, D1, and D2.

\section{\phx Results} \label{sec:results}

\begin{figure}[ht]
\centering
\gridline{\fig{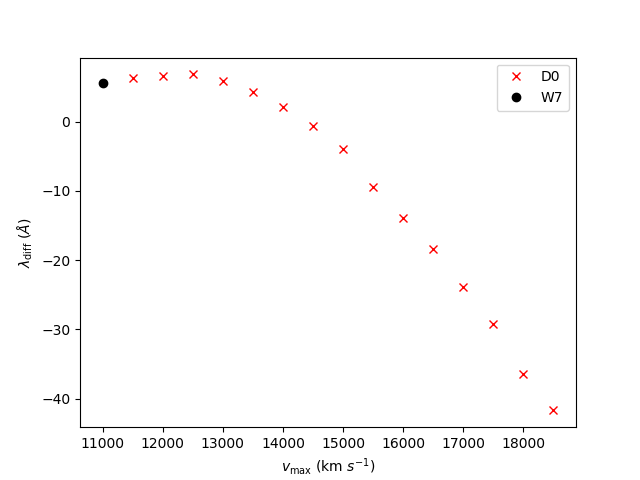}{0.45\textwidth}{(a)}
}
\gridline{\fig{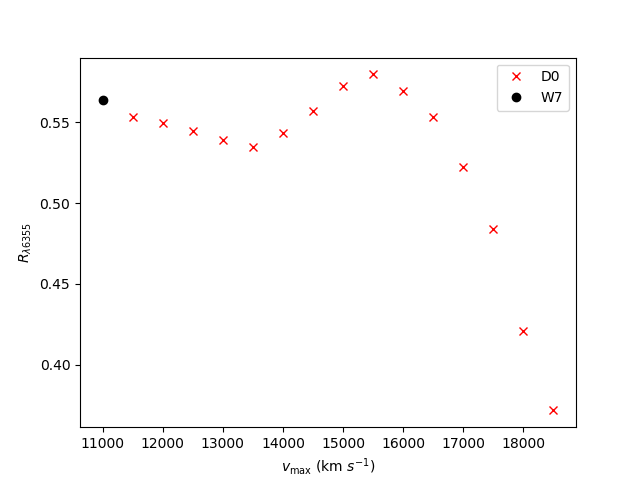}{0.45\textwidth}{(b)}
}
\gridline{\fig{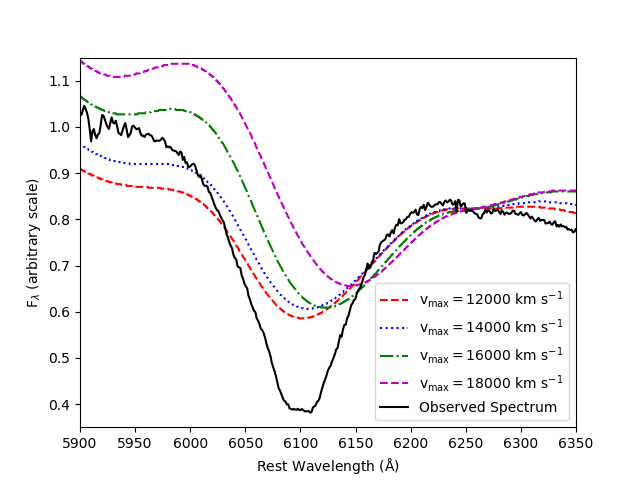}{0.45\textwidth}{(c)}
}
\caption{$\lambda_{\text{diff}}$ (a) and $R_{\lambda 6355}$ (b) vs. $v_{\text{max}}$ (for D0 fits) at day $+1$ with $v_{\text{min}}$ and $v_{e}$ held fixed at $9000$~\kmps and $1000$~\kmps, respectively.  $\lambda_{\text{diff}}$ trends monotonically downward, while $R_{\lambda 6355}$ peaks at $v_{\text{max}} = 15500$~\kmps.  The values of $\lambda_{\text{diff}}$ and $R_{\lambda 6355}$ for unmodified W7 are included for comparison.
(c): Fits to the \ion{Si}{2} $\lambda 6355$ line for several
  values of $v_{\text{max}}$. Note that the lineshape is best fit with $v_{\text{max}}$
  between $14000$~\kmps and $16000$~\kmps, but the line is pushed too far to the red as $v_{\text{max}}$ increases.   }
\label{fig17}
\end{figure}

\begin{figure}[ht]
\centering
\gridline{
\fig{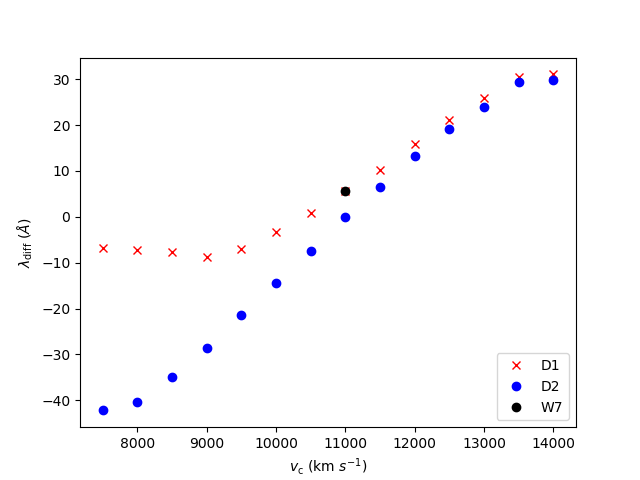}{0.45\textwidth}{(a)}
\fig{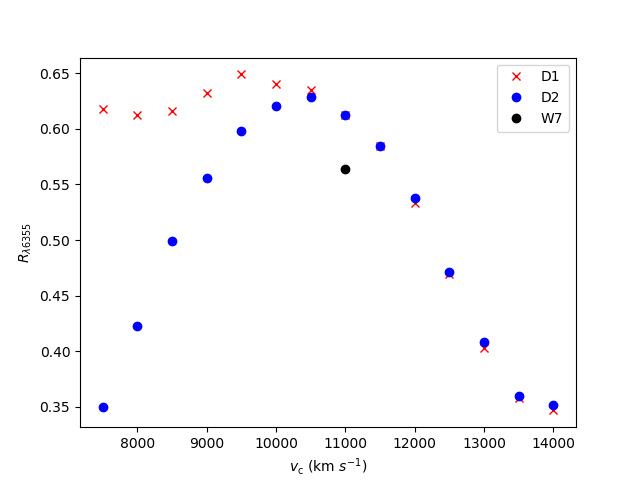}{0.45\textwidth}{(b)}
}
\gridline{
  \fig{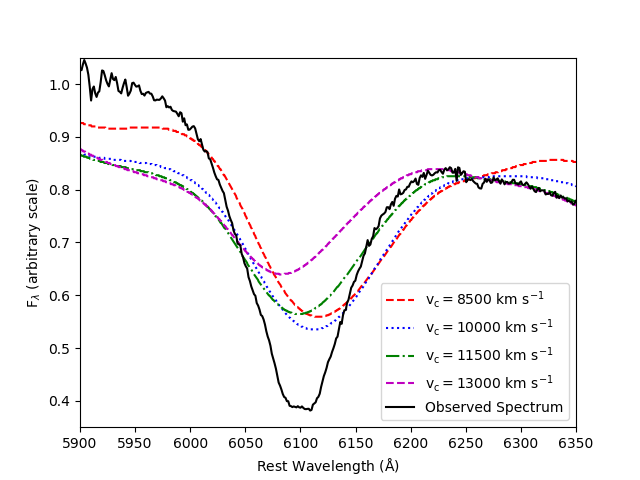}{0.45\textwidth}{(c)}
\fig{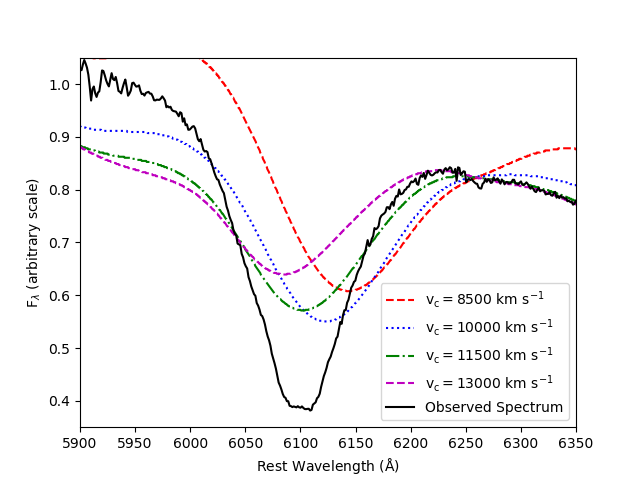}{0.45\textwidth}{(d)}
}
\caption{$\lambda_{\text{diff}}$ (a) and $R_{\lambda 6355}$ (b) vs. $v_{c}$ (for D1 and D2 fits) at day $+1$ with w and p held fixed at $3000$~\kmps and $2$, respectively. In the D2 models, $v_{u}$ is held fixed at $17000$~\kmps. In D1, $\lambda_{\text{diff}}$ increases monotonically and $R_{\lambda 6355}$ peaks at $9500$~\kmps.  In D2, $\lambda_{\text{diff}}$ increases monotonically and $R_{\lambda 6355}$ peaks at $10500$~\kmps.  The values of $\lambda_{\text{diff}}$ and $R_{\lambda 6355}$ for unmodified W7 are included for comparison. (c): D1 fits to the \ion{Si}{2} $\lambda 6355$ line for several values of $v_{c}$.  (d): D2 fits to the \ion{Si}{2} $\lambda 6355$ line for several values of $v_{c}$.}
\label{fig18}
\end{figure}

We converged grids of W7 models at day $+1$, $+4$, $+8$, $+12$, and $+18$ in D0, D1, and D2. 
Luminosities were chosen for each phase based on the observed luminosity evolution of SN~2012fr. 
The goodness-of-fit of the \ion{Si}{2} $\lambda 6355$ line was measured using $\lambda_{\text{diff}}$ and $R_{\lambda 6355}$.  
To compare the fixed-time effects of each procedure, we used the day $+1$ fits, which were generally the best.  
While all the input parameters affect the fits, the most significant for us are those which control the location of the modification in velocity space.  
In D0, we focused on the effect of changes in $v_{\text{max}}$ and in D1 and D2 we focused on $v_{c}$.  
We did this because $v_{\text{max}}$ determines the extent and prominence of the steepened region of the density profile in D0 and $v_{c}$ determines the location of the shell in D1 and D2.  
Thus, these parameters are most directly related to where mass is being added or subtracted in velocity space.  
Selected plots of $\lambda_{\text{diff}}$ and $R_{\lambda 6355}$ against velocity parameters of D0, D1, and D2 are shown in Figures~\ref{fig17} and ~\ref{fig18}.  \\

As shown in Figure~\ref{fig17}, increasing $v_{\text{max}}$ with $v_{\text{min}}$ and $v_{e}$ held fixed caused $\lambda_{\text{diff}}$ to decrease.  
This is not surprising.  A steeper density profile would shift the mean of the \ion{Si}{2} distribution to a lower velocity.  
More interesting is the sharp peak in $R_{\lambda 6355}$ at $15500$~\kmps, which indicates a maximum in absorption strength.  
The peak may be the result of weakening \ion{Si}{2} $\lambda 5792$ absorption raising the flux at $5970 $~\ang, thus increasing the area under the SPC.   
The sharp decline in flux for values above $15500$~\kmps is probably due to decreasing \ion{Si}{2} opacity as more mass is removed from the $9000 - 16000$~\kmps region and the corresponding red-ward movement of the absorption line.  
What this suggests is that the shape of the \ion{Si}{2} $\lambda 6355$ line is slightly better fit by a density profile that falls off more quickly in the range where \ion{Si}{2} is found than it does in unmodified W7.  \\

As shown in Figure~\ref{fig18}, increasing $v_{c}$ in D1 produces faster \ion{Si}{2} absorption, as expected.  
Shifting a shell to higher speeds increases the average velocity of \ion{Si}{2}.  
However, $R_{\lambda 6355}$ begins to fall off starting at $v_{c} = 11000$~\kmps.  
This may be due to the broadening of the line as optically thick \ion{Si}{2} is spread over a wider range of velocities.  
It may also be due to the strengthening of the \ion{Si}{2} $\lambda 5972$ line, or an increase in the mass fraction of the HV \ion{Fe}{2} mentioned earlier.  
To probe the significance of the HV Fe, we converted the $^{52} \text{Fe}$, $^{54} \text{Fe}$ and $^{56} \text{Fe}$ abundances in W7 to Si and re-converged the model. 
We found that the removal of HV Fe had a negligible effect on the formation of the $5500-6500$~\ang region of the spectrum, suggesting that the strong absorption may be produced by the LV Fe that forms the continuum.  
Previous studies \citep[such as][]{bongard08} have shown that Fe absorption at low velocities significantly affects the shape of the continuum in certain wavelength regions at late time, including the region considered here.  
  
In any case, we find that shells centered at high velocities ($\geq 11000$~\kmps) did not produce good fits.  
Better fits are obtained with the shell centered at $\leq 11000$~\kmps and these produce stronger absorption profiles than the best fits using D0.
Therefore, our preliminary analysis suggests that a shell centered somewhere below $11000$~\kmps can account for SN~2012fr better than an unusually steep density profile.

\begin{figure}[ht]
\centering
\includegraphics[scale=1]{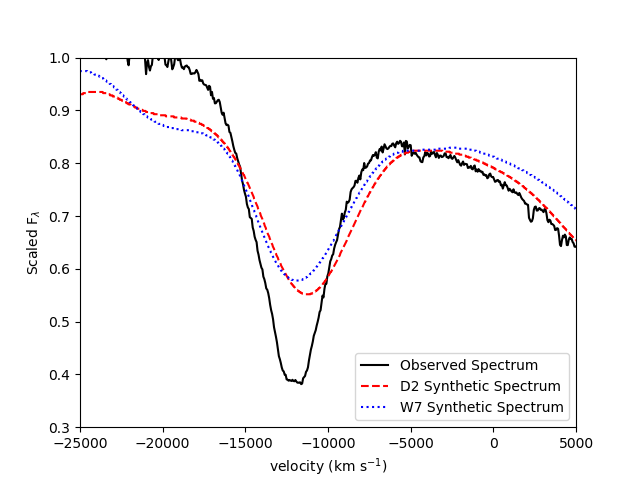}
\caption{D2 fit (red) to the \ion{Si}{2} $\lambda 6355$ line in SN~2012fr at day $+1$, with unmodified W7 fit (blue) for comparison, plotted in velocity space (negative velocities are indicate blue-shift).  The profiles have nearly the same shape, the W7 profile being slightly too blue and the D2 profile being somewhat too red.  Neither profile successfully captures the flat bottom (doublet feature) of the observed line.  At this epoch, the D2 absorption trough is slightly deeper, but the wavelength minimum of W7 is closer to observations.  }
\label{fig20}
\end{figure}

\begin{figure}[ht]
\centering
\gridline{
\fig{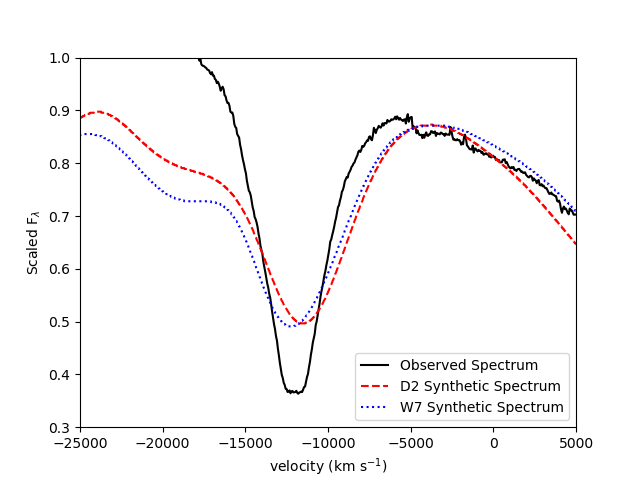}{0.45\textwidth}{(a)}
\fig{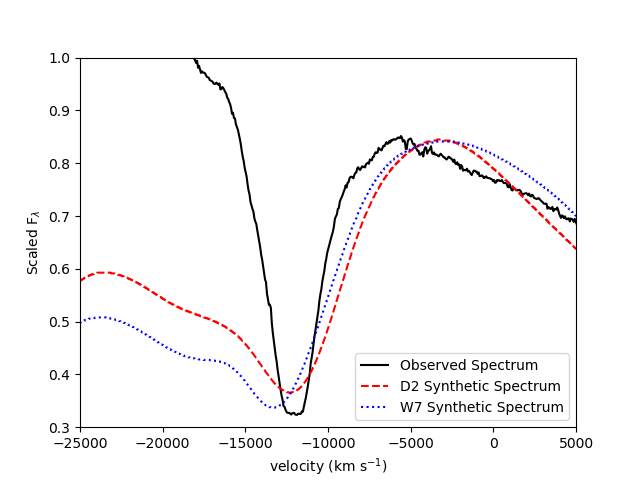}{0.45\textwidth}{(b)}
}
\caption{D2 fit at day $+4$ (left) and day $+8$ (right) with unmodified W7 for comparison.  }
\label{fig21}
\end{figure}

\begin{figure}[ht]
\centering
\includegraphics[scale=0.65]{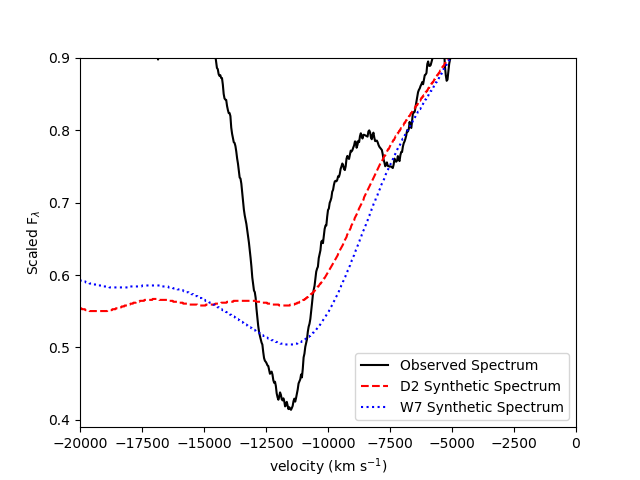}
\caption{D2 fit at day $+18$ (right) with unmodified W7 for comparison.  }
\label{fig22}
\end{figure}

Increasing $v_{c}$ in D2 (Figure~\ref{fig18}) produces a more interesting effect. $\lambda_{\text{diff}}$ increases linearly, while $R_{\lambda 6355}$ peaks near $v_{c} =10500$~\kmps at a value of $0.628$. 
Moreover, $\lambda_{\text{diff}} = -8$~\ang at that point, so the position is reasonably well-fit (and improves in later fits).  
Figure~\ref{fig20} shows the day $+1$ D2 fit with unmodified W7 in blue for comparison.  
D2 parameters used are $v_{c} = 10500$~\kmps, w = $3000$~\kmps, p = $2$, and $v_{u} = 17000$~\kmps.  
Although the D1 fit at day $+1$ is actually slightly worse than W7, the overall D2 time series for these parameters does a better job of matching observations than W7.  
Figure~\ref{fig21} shows the fits at day $+4$ (left) and day $+8$ (right) using the same parameters.  
In both fits, the synthetic line is centered correctly and the absorption profile is a slightly better match than in the unmodified spectrum.  
Even at later times, when W7 ceases to resemble the observed spectra of SNe Ia in the $5500 - 6500 $~\ang wavelength region \citep{bbbh06}, the situation is still improved qualitatively.  
In Figure~\ref{fig22}, the fit at day $+18$ (right) is shown for the same D2 parameters.  
At day $+18$, the \ion{Si}{2} line is completely washed out by Fe features in the unmodified fit, but in the D2 fit faint \ion{Si}{2} absorption remains distinct and correctly positioned.

It may be that constraining the \ion{Si}{2} abundance in W7 to occupy a smaller range of velocities would produce a better result.  
We reemphasize that the procedures used here do not affect the velocities at which \ion{Si}{2} is found or the mass fractions in each velocity bin, since they are density enhancements rather than abundance enhancements.  
By only modifying the density, we consider the effect of
redistributing mass without changing the velocity intervals in which
each ion is present. As mentioned earlier, \autoref{fig:si_extent} indeed shows that the
  velocity extent of silicon in W7 well matches the velocity extent
  that we have inferred from our SYNOW fits. Due to the different
  treatments of radiative transfer, there is a global velocity offset
  between SYNOW and PHOENIX,
  but the velocity width is nearly the same.
We modified only the density here because modifying the abundance
distribution as well would produce a much larger parameter space.   
We do, of course, alter the optical depths at every point, both the
Sobolev optical depths, which are not used in \phx, and in
general. Since \phx does not use Sobolev optical depths the line
formation in \phx is much more complex than in SYNOW \citep{bongard08}.  We note that since we ran over $1000$ models in this study, we ran them in Local Thermal Equilibrium (LTE) for computational expediency.  However, NLTE effects are known to play an important role in the formation of the features being studied here \citep{bbbh06}

\section{Discussion} \label{sec:discussion}

From the results presented in \S~\ref{sec:synow}, it is evident that the abundances in SN~2012fr are highly stratified at early and late times.  
HV components of \ion{Si}{2} and \ion{Ca}{2} occupy roughly the same velocity region at early times, possibly mixed with \ion{Mg}{2}.  However, at later times PV \ion{Ca}{2} and PV \ion{Si}{2} diverge, \ion{Si}{2} detaching from the photosphere before maximum light and \ion{Ca}{2} remaining attached (again, as constrained by the H\&K feature). 
The narrowness of the \ion{Si}{2} $\lambda 6355$ line, its shallow velocity evolution and its prominence at late times further complicate the situation.  
The question that we attempted to answer in \S~\ref{sec:density} - \S~\ref{sec:results} is whether this behavior can be explained by only an abnormal density structure.    

Our analysis using W7 suggests that it might be possible.  
The D1 and D2 procedures both produced slightly improved fits to the \ion{Si}{2} $\lambda 6355$ line after maximum light.   
Both procedures produced the best results when shells were centered between $10000$~\kmps and $11000$~\kmps.  
However, shells at higher velocities produced much worse fits, as evidenced by the low values of $R_{\lambda 6355}$ obtained for large values of $v_c$ in D1 and D2.  
What this shows is that if there is a narrow shell in SN~2012fr, it is likely centered at or below $11000$~\kmps in velocity space.
D0, on the other hand, does not improve the fits overall.  
While the shape of the P-Cygni profile does improve slightly in D0 (see Figure~\ref{fig12}), the line is formed at too low a velocity.  

Part of the difficulty of interpreting these results has to do with the W7 density structure itself.  
In W7, flame quenching produces a sharp density spike near
$15000$~\kmps \citep{bbbh06} that is clearly visible in
Figures~\ref{fig12}, \ref{fig13}, and~\ref{fig14}.  
This feature is not removed in any of our procedures.  
Instead, added features are multiplicatively superimposed onto local features to create a composite structure.  In D0, this means the spike is ``dampened'', somewhat, while it may be enhanced in D1 and D2 based on where the shell is positioned.  
The reason for doing this is to ensure that the differences between modified and unmodified W7 depend (as much as possible) only on the parameterization of the procedure rather than on local variability in the W7 profile. 
The disadvantage is that W7 may already contain the feature being tested for, and so by failing to remove the bump, the added density enhancements at high velocities might be much too strong.

It is unclear whether a low-velocity shell is implied by our results.  
As shown in Figures~\ref{fig20}--\ref{fig22}, D2 reproduced the temporal gradient of \ion{Si}{2} $\lambda 6355$ in SN~2012fr reasonably well.  
The best fits were obtained when the D2 shell was placed so as to add mass to the lower $3000$~\kmps or so of the \ion{Si}{2} distribution and subtract mass from the rest. 
If, as suggested by SYNOW, most of the \ion{Si}{2} in SN~2012fr is confined to a velocity interval about $2000-3000$~\kmps across, then this structure makes sense.   
If PV \ion{Ca}{2} is present throughout this shell, then the distinctiveness of the Ca IR3 lines at late times is not surprising.  
Placing more \ion{Ca}{2} within this range would reduce the relative spread
of the \ion{Ca}{2} abundance, causing the lines to become narrow.     
The same can be said of the distinct separation of the \ion{Fe}{2} lines that form in the $4500 - 5000$~\ang range, assuming any \ion{Fe}{2} is present that high.  

Furthermore, if there is a shell, it is fairly clear where it has to be located.  
To have a significant effect on the W7 \ion{Si}{2} distribution, shells have to be placed at high enough speeds to interact with the $9000 - 16000$~\kmps region.  
However, as demonstrated, they produce low-quality fits above $11000$~\kmps.  
So, if SN~2012fr has a shell, it is probably centered somewhere between $9000 - 11000$~\kmps.  
Moreover, it needs to be a thick shell to maintain high opacity at late times and keep the \ion{Si}{2} $\lambda 6355$ line distinct from Fe features.  
The D1 shell displayed in Figure~\ref{fig13} increases the unnormalized mass of the SN by roughly $24$\%.  
The amount of mass added varies according to the input parameters, but in any case it must be large to significantly affect the spectra.  
After normalization, the density profile flattens out significantly, producing a much shallower density gradient.  
That late-time behavior, illustrated by our day $+18$ fit with D2 (Figure~\ref{fig22}) seems to suggest that the a shell is capable of keeping the \ion{Si}{2} feature distinct from Fe features later than usual, although this conclusion is somewhat provisional because of the very poor fit quality at day $+18$ in W7.  

A clearer picture of SN~2012fr can likely be gained by a procedure to
vary the  abundances, particularly those of Si and Ca, in W7 or some other model.  
It may be that the peculiar features of SN~2012fr can be explained
entirely in terms of how these elements are distributed, without any
appeal to the global density structure, if there were such a thing as
a fiducial SN Ia model.  
This is what SYNOW does on a qualitative level, but not to a degree of physical accuracy to validate a concrete conclusion. 
\citet{childress13b} suggested that the odd behavior of the \ion{Si}{2} $\lambda 6355$ line could be explained by a cutoff in the radial distribution of IMEs at a higher-than-usual velocity.  
Such a model could be tested by converting Si to Fe or $^{56} \text{Ni}$ below some minimum velocity (perhaps $11000$~\kmps), and would not require any enhancement of the density profile.  
Or, if there is a shell, a correct model of it would have to take both the density and abundance structures into account.  
Such an analysis is beyond the scope of this work.

\section{Conclusions} \label{sec:conclusion}

We demonstrated here using various modeling techniques that the unusual spectroscopic features of SN~2012fr can be described in a sensible way.  
Using SYNOW, we found that \ion{Ca}{2} and \ion{Si}{2} each have two component (HV and PV)  that behave similarly up to maximum light.  
After maximum, the minimum velocities of the PV components diverge, \ion{Ca}{2} remaining attached to the photosphere and \ion{Si}{2} detaching with a constant minimum velocity of $14000$~\kmps.  
Using W7 converged in~\phx, we probed the density structure of SN~2012fr using a set of parameterized density enhancement procedures.  
We focused our analysis on the \ion{Si}{2} $\lambda 6355$ line and developed goodness-of-fit parameters $\lambda_{\text{diff}}$ and $R_{\lambda 6355}$ to aid in quantitative analysis.  Using these parameters, we ran grids in each parameter space and analyzed the effect of each procedure.  
We found that the position, absorption strength, and temporal evolution of $\lambda 6355$ in SN~2012fr can be modeled with slightly improved accuracy using a density enhancement centered at low velocities.   
The best results were obtained for profiles that contained density shells centered between $9000$ and $11000$~\kmps.  
We also hypothesized that the presence of thick \ion{Ca}{2} in the shell region would explain the clean separation of the Ca IR3 lines in SN~2012fr.  
Within the Delayed Detonation paradigm the formation of a shell is
rather natural. The initial deflagration provides a certain amount of
energy to the WD, whether the energy is sufficient to unbind the WD
depends on the parameters (in the 1-D DDT approximation). If the
energy is not quite enough to unbind the WD, material begins to fall
back, which creates a shell when hit by the shock produced by the
detonation. From our density plots, this corresponds to a DDT
transition density of $\sim 10^7$~\gcm, which is the right order of
magnitude. Thus, 
future research will explore the effect of density and abundance
modifications using a Delayed Detonation (DDT) model converged in \phx.
This analysis will include consideration variations of the transition density
$\rho_{tr}$ \citep{H02,contreras12fr18}) at which the deflagration
becomes a detonation, and how the placement of $\rho_{tr}$ affects the abundance structure of the model.  

\acknowledgments
Support from NSF grants AST 1008343, AST 1613426, AST 1613455, and AST
1613472 is gratefully acknowledged. 
EB and CC acknowledge partial support from NASA Grant NNX16AB25G. EB
thanks  the Aarhus University Research Fund (AUFF) for a Guest
Researcher grant. 
CC gratefully acknowledges support from NSF REU Grant PHY1659501.
M. D. Stritzinger is supported by a research grant 13261 (PI Stritzinger) from the Villum FONDEN and  is also grateful to Aarhus University's Faculty of Science \& Technology for a generous sabbatical grant. 
This research used resources of the
National Energy Research Scientific Computing Center (NERSC), which is
supported by the Office of Science of the U.S.  Department of Energy
under Contract No.  DE-AC02-05CH11231; and the H\"ochstleistungs
Rechenzentrum Nord (HLRN).  We thank both these institutions for a
generous allocation of computer time.

\bibliography{refs}

\clearpage
\renewcommand{\thefigure}{A\arabic{figure}}
\setcounter{figure}{0}

\appendix
\section{SYNOW Fits}
\label{sec:appx}

\begin{figure}[ht]
\centering
\gridline{
\fig{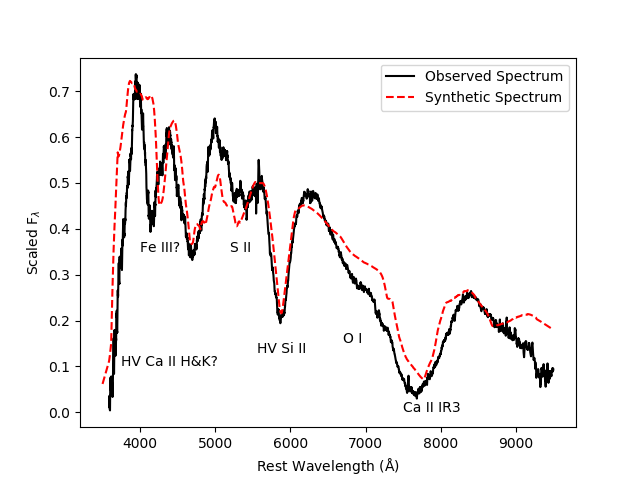}{0.45\textwidth}{(a)}
\fig{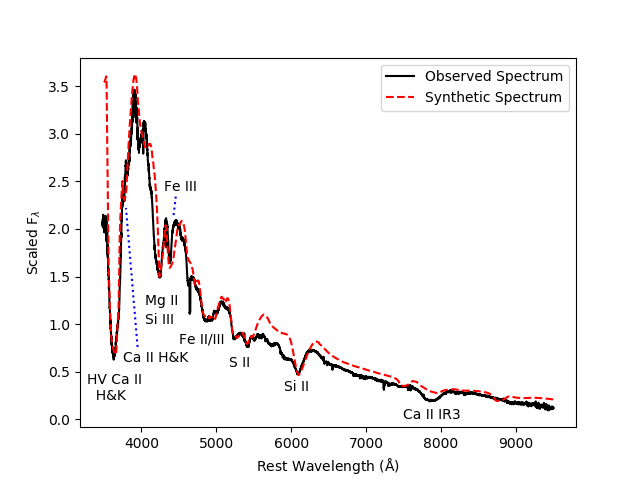}{0.45\textwidth}{(b)}
}
\gridline{
\fig{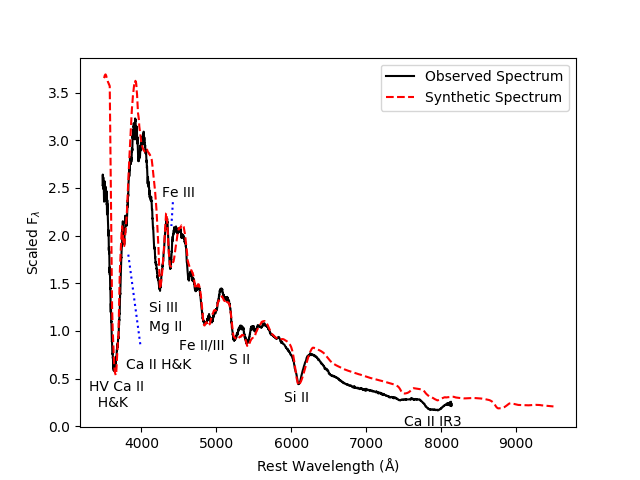}{0.45\textwidth}{(c)}
\fig{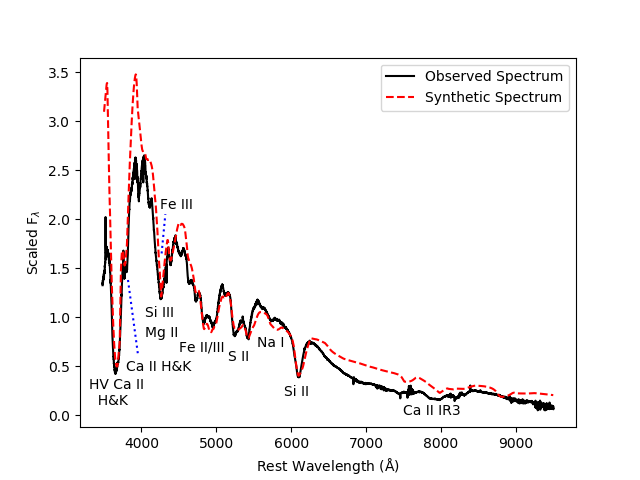}{0.45\textwidth}{(d)}
}
\caption{Premaximum spectra with SYNOW fits. (a): Day $-15$ spectrum and a SYNOW fit.  The fit is very good in
  the $5500 - 6500 $~\ang range and is reasonable in region where Ca
  IR3 forms, but is not good blue-ward of $5500$~\ang.  \ion{O}{1} is
  tentatively identified (though not included in Table~\ref{tab1})
  along with \ion{Mg}{2} to account for some of the absorption
  red-ward of \ion{Si}{2} $\lambda 6355$. (b): Day $-8$ spectrum and a
  SYNOW fit. (c): Day $-5$ spectrum and a
  SYNOW fit. (d): Day $-2$ spectrum and a SYNOW fit.\label{fig:premax}}
\end{figure}

\autoref{fig:premax} shows the premaximum observed spectra and SYNOW fits
at days $-12$, $-8$, $-5$, and $-2$.

\begin{figure}[ht]
\centering
\gridline{
\fig{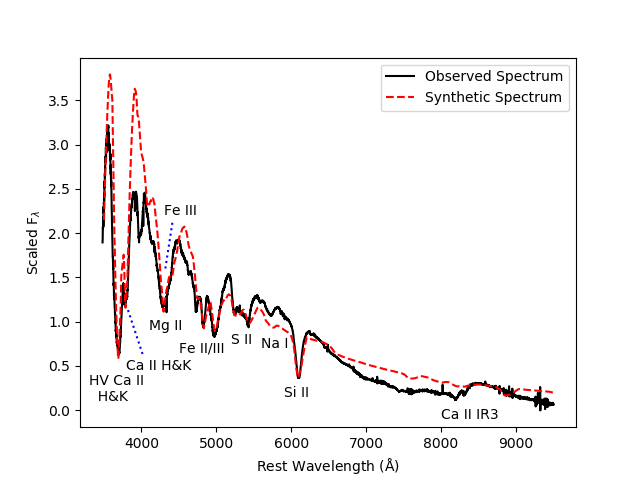}{0.45\textwidth}{(a)}
\fig{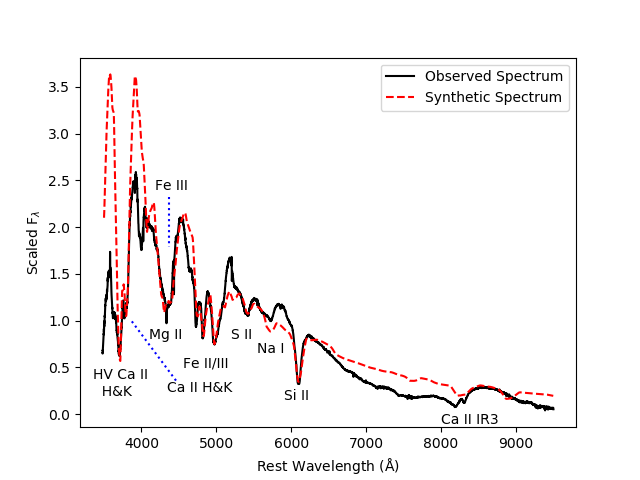}{0.45\textwidth}{(b)}
}
\gridline{
\fig{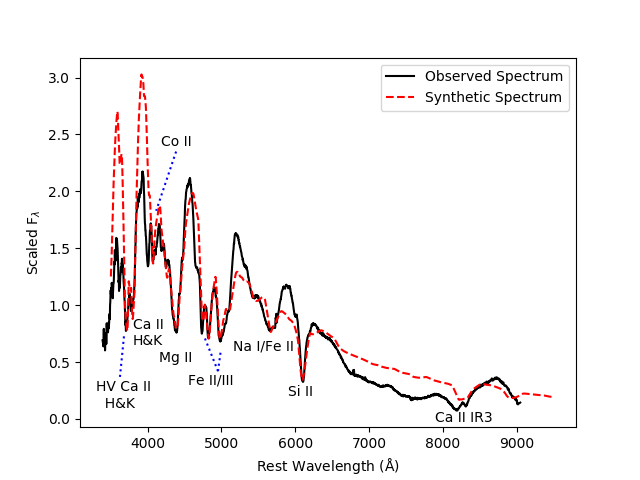}{0.45\textwidth}{(c)}
\fig{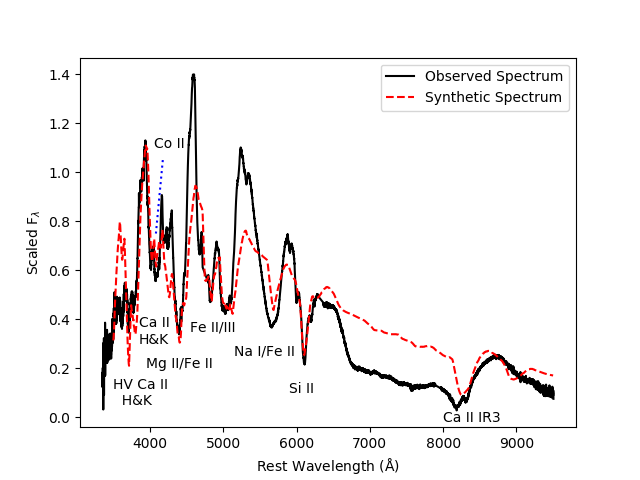}{0.45\textwidth}{(d)}
}
\caption{(a): Day $+4$ spectrum and a SYNOW fit. (b): Day $+8$
  spectrum and a synthetic SYNOW fit. (c) Day $+12$ spectrum and a
  SYNOW fit. (d): Day $+18$ spectrum and a SYNOW fit.}
\label{fig8}
\end{figure}

\autoref{fig8} shows the postmaximum observed spectra and SYNOW fits
at days $+4$, $+8$, $+12$, and $+18$.

\begin{figure}[ht]
\centering
\fig{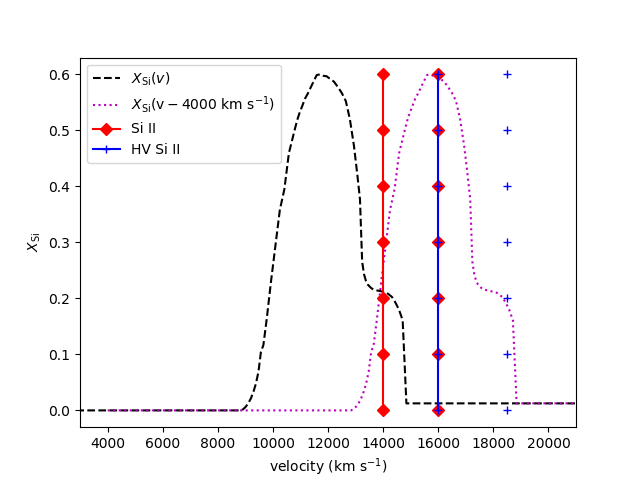}{0.8\textwidth}{(a)}
\caption{Comparison of the \ion{Si}{2} distribution in W7 compared to the minimum and maximum PV and HV ion velocities in the day $+1$ SYNOW fit.  When the W7 mass fraction function is shifted to the right by $4000$~\kmps, the distributions agree quite nicely.  The red and blue markers denote min/max SYNOW velocities as in Figure~\ref{vel_fig}}
\label{fig:si_extent}
\end{figure}

\autoref{fig:si_extent} 
  shows the velocity extent of silicon in W7 compared to 
  that we have inferred from our SYNOW fits. Due to the different
  treatments of radiative transfer, there is a global velocity offset
  between SYNOW and PHOENIX,
  but the velocity width is nearly the same.

\end{document}